\documentclass[twocolumn,english,prl,longbibliography,superscriptaddress]{revtex4-2}
\usepackage[T1]{fontenc}
\usepackage[latin9]{inputenc}
\usepackage{color}
\usepackage{babel}
\usepackage{float}
\usepackage{amsmath}
\usepackage{amssymb}
\usepackage{graphicx}
\usepackage[unicode=true,
 bookmarks=false,
 breaklinks=true,pdfborder={0 0 1},colorlinks=true]
 {hyperref}
\hypersetup{
 citecolor=blue,linkcolor=blue,urlcolor=blue}

\newcommand{\msf}[1]{\mathsf{#1}}	



\begin{document}
\title{Tunable viscosity across the BCS-BEC crossover}

\author{Yunxiang Liao}
\affiliation{Department of Physics, KTH Royal Institute of Technology, SE-106 91
Stockholm, Sweden}

\author{Andrey Grankin}
\affiliation{Joint Quantum Institute, Department of Physics, University of Maryland,
College Park, MD 20742, USA}

\author{Archisman Panigrahi}
\affiliation{Massachusetts Institute of Technology, Cambridge, MA 02139, USA}

\author{Victor Galitski}
\affiliation{Joint Quantum Institute, Department of Physics, University of Maryland,
College Park, MD 20742, USA}

\author{Leonid Levitov}
\affiliation{Massachusetts Institute of Technology, Cambridge, MA 02139, USA}

\begin{abstract}
Tunable interactions  make ultracold quantum gases a unique platform for exploring  hydrodynamic properties in the strongly correlated regime. Of particular interest are turbulent flows possible in the regime of high Reynolds numbers. Since the system size and flow velocity are limited in experimentally realistic systems, we propose an alternative approach to enhance the Reynolds numbers in an ultracold Fermi gas by minimizing the shear viscosity in the vicinity of the Feshbach resonance. By employing the Keldysh formulation of the linear response theory, we theoretically demonstrate that the shear viscosity can vary by several orders of magnitude in the vicinity of the BCS-BEC crossover. It is also shown that while Drude-like contributions generally dominate at large Feshbach detunings, higher-order vertex corrections, including the Maki-Thompson contribution, become significant and suppress singular behavior in the near-resonant regime. Our results provide a roadmap for achieving tunable Reynolds numbers in ultracold quantum fluids, which can serve as table-top turbulence simulators.
\end{abstract}
\maketitle
Ultracold quantum gases \cite{bloch2008many, ketterle2008making,giorgini2008theory} provide a rich platform for exploring the interplay between microscopic interactions and macroscopic hydrodynamics \cite{joseph2007measurement}, enabled by the tunable interaction strength and confinement properties.
By utilizing Feshbach resonance \cite{chin2010feshbach}, the interaction strength can be
adjusted continuously to sweep between a molecular Bose-Einstein
condensate (BEC) and fermionic Bardeen-Cooper-Schrieffer (BCS) superfluid
in a way impossible for conventional fluids \cite{regal2004observation}. Experimentally the hydrodynamic properties of cold gases can be probed by various means including collective
oscillation dynamics \citep{patel2020universal} or the cloud expansion properties
\citep{cao2011universal,Exp-1,Exp-2}. 

\begin{figure}[t]
\begin{centering}
\includegraphics[width=1\columnwidth]{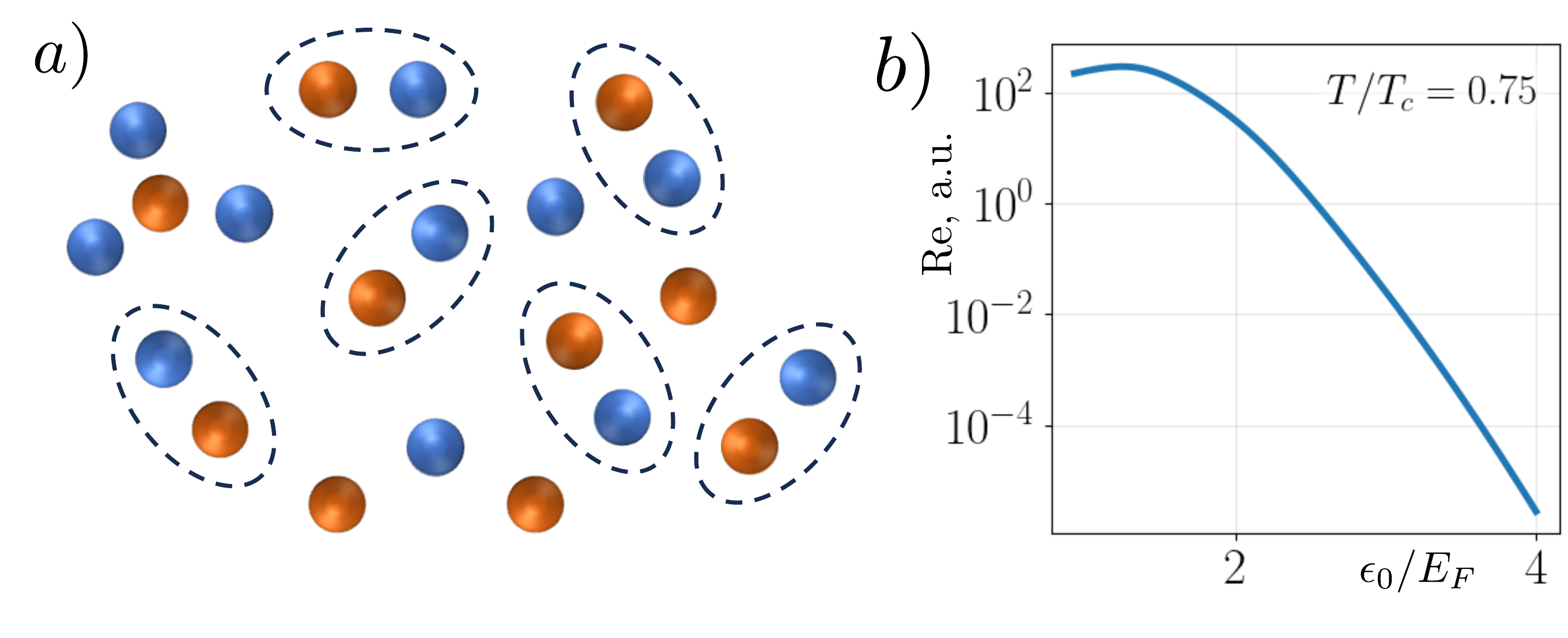} 
\par\end{centering}
\caption{Viscosity in the two-channel model across BCS-BEC crossover. (a) Schematic illustration of fermion pairs forming bosonic molecules.  (b) Reynolds number, or inverse shear viscosity (in arbitrary units) as function of detuning $\epsilon_0$ at $T=0.75 T_c$, where $T_c$ is the global BEC transition temperature. The Reynolds number is expected to change over several orders of magnitude.}
\label{Fig0} 
\end{figure}

The hydrodynamic processes are governed by the bulk $\zeta$ and shear viscosities $\eta$ of the gas, which quantify the degree of ``fluid friction'' against deformation or shear. Shear viscosity $\eta$ quantifies a fluid's resistance to shearing
flow, defined as the motion of adjacent layers relative to one another
\citep{landau1987fluid}. Fundamentally, it controls the onset of
turbulence, a regime governed by the competition between inertial
forces and viscous damping \cite{landau1987fluid}, parametrized by the Reynolds number
$\text{Re}=u\rho L/\eta$, where $u$ is the characteristic flow velocity,
$L$ is the characteristic length-scale of the flow and $\rho$ is the
particle density. However, reaching the high-Re turbulent regime \cite{schafer2009nearly} remains
a challenge in ultracold systems due to constraints on system size
and  flow velocities. Minimizing the shear viscosity~\cite{Fluct} thus
represents the most viable way to achieve turbulence. As a transport
coefficient, $\eta$ is highly sensitive to the interaction strength,
and therefore, the use of Feshbach resonance to tune both
makes ultracold atoms a perfect table-top  turbulence simulator.

Theoretically, both shear and bulk viscosities were previously studied using BCS-type single-channel model \cite{kagamihara2019shear, enss2011viscosity}. These studies \cite{kagamihara2019shear} demonstrated that the viscosity minimum is located slightly away
from unitarity. In this work, we examine the viscosity using a two-channel
model, which, unlike its single-channel counterpart, explicitly includes
a resonant state. This approach has several advantages \citep{gurarie2007resonantly}.
First, it enables exploring the interplay between molecular bound  states in the closed channel and scattering states of free fermions in the open channel.
In particular, the two-channel model retains the closed-channel molecular degree of freedom explicitly, and as a result provides a description applicable to both  wide and narrow Feshbach resonances.
In contrast, the single-channel model integrates out the closed-channel molecular degree of freedom, and can be viewed as the wide resonance limit of the two-channel model.
Unlike the two channel model, it does not naturally resolve distinct fermionic, bosonic, and mixed contributions to the viscosity within the same framework.
Furthermore, in the narrow resonance limit, the two-channel model introduces a small parameter, the atom-molecule coupling constant $g$, which determines the width of the resonance~\citep{gurarie2007resonantly}, enabling a perturbatively controlled and physically transparent treatment of transport.
We utilize this property of the two-channel model in combination with linear response theory \cite{taylor2010viscosity} to derive the shear viscosity  perturbatively.
We find that by tuning
the Feshbach resonance, the shear viscosity and therefore the Reynolds number can be changed by several
orders of magnitude, as shown in Fig~\ref{Fig0}(b). Furthermore, in our work, we employ the Keldysh 
formalism that allows us to avoid issues with analytic continuation, which leads to various complications
as discussed in \cite{kagamihara2019shear}. 

We start by defining the Hamiltonian of the two-channel model  $H=H_{0}+H_{\text{int}}$, representing fermions binding into bosonic pairs as schematically shown in Fig.~\ref{Fig0}~(a):
\begin{align}
H_{0} & =\sum_{{\bf k},\sigma}\xi_{{\bf k}}^{F}\psi_{\sigma}^{\dagger}({\bf k})\psi_{\sigma}({\bf k})+\sum_{{\bf q}}\xi_{{\bf q}}^{B}\phi^{\dagger}({\bf q})\phi({\bf q}),\label{eq:H0}\\
H_{\text{int}} & =\frac{g}{\sqrt{V}}\sum_{{\bf q},{\bf k}}\left\{ \phi({\bf q})\psi_{\uparrow}^{\dagger}({\bf q-k})\psi_{\downarrow}^{\dagger}({\bf k})+\text{H.c.}\right\} .\label{eq:Hint}
\end{align}
Here $\xi_{{\bf k}}^{F}=k^{2}/2m-\mu,$ $\xi_{{\bf q}}^{B}=q^{2}/4m-\left(2\mu-\epsilon_{0}\right)$
are the fermionic and bosonic dispersions respectively $\epsilon_{0}$
denotes the detuning, $m$ is the fermionic mass and $\mu$ is the
chemical potential. $\psi_{\sigma}$ $(\psi_{\sigma}^{\dagger})$
and $\phi$ $\left(\phi^{\dagger}\right)$ are the annihilation (creation)
operators of fermions and bosons respectively. $g$ denotes the coupling
strength and $V$ is the volume of the system. In addition, we impose
the conservation of the total number of atoms.

Our goal is now to find the shear viscosity in the two-channel model.
To this end, we derive expressions for the stress tensor from the
conservation of the momentum densities $g_{i}$ along direction $i$ : $\partial_{t}g_{i}+\partial_{j}T_{ij}=0$.
As shown in the Supplementary Material~\cite{Sup}, the total stress-tensor
$T_{ij}$ can be written as a sum of the pure bosonic and fermionic contributions $T^{(B)}_{ij}$ and $T^{(F)}_{ij}$, as well as an interaction term $T_{ij}^{(\msf{int})}$ that does not contribute to the off-diagonal components ($i\neq j$):
						\begin{subequations}
                        \begin{align}
								& T^{(F)}_{ij} =
								\frac{1}{2m}
								\sum_{\sigma}
								\left[ 
								\partial_i \psi^{\dagger}_{\sigma} 
								\partial_j \psi_{\sigma} 
								+
								\partial_j \psi^{\dagger}_{\sigma} 
								\partial_i \psi_{\sigma} 
								-\frac{1}{2}
								\delta_{ij} \nabla^2 \left( \psi^{\dagger}_{\sigma} \psi_{\sigma} \right) 
								\right],
								\label{eq:TF}
								\\
								&T^{(B)}_{ij}
								=\,
								\frac{1}{4m}
								\left[ 
								\partial_i \phi^{\dagger} 
								\partial_j \phi 
								+
								\partial_j \phi^{\dagger} 
								\partial_i \phi
								-\frac{1}{2}
								\delta_{ij} \nabla^2 \left( \phi^{\dagger} 
								\phi\right) 
								\right],
								\label{eq:TB}
                                \\
								&T_{ij}^{(\msf{int})}
								=
								\frac{g}{2}	
								\delta_{ij}
								\left[
								\phi^{\dagger}
								\psi_{\downarrow}  
								\psi_{\uparrow} 
								+
								\phi
								\psi^{\dagger}_{\uparrow} 
								\psi_{\downarrow}^{\dagger}  
								\right].
								\label{eq:Tint}
							\end{align}	
						\end{subequations}         
We emphasize that the shear viscosity is determined by the off-diagonal stress tensor $T_{xy}$, which receives contributions only from the non-interacting components, $T_{xy}=T_{xy}^{(F)}+T_{xy}^{(B)}$.


Throughout this work, we employ the weak-coupling limit $g\rightarrow0$
(or the so-called narrow resonance regime) and evaluate the viscosity
perturbatively using linear response theory. This approach corresponds
to driving the system with a weak perturbation proportional to the
total stress tensor $T_{xy}$ and then evaluating
the average $\left\langle T_{xy}\right\rangle $ in the driven state.
In this case, the total shear viscosity can be expressed using the Kubo
formula as follows \citep{PhysRevA.81.053610}:

\begin{equation}
\eta=\lim_{\omega\rightarrow0,q\rightarrow0}\frac{1}{\omega}\text{Im}\left\{ i\int_{0}^{\infty}e^{i\omega t}\left\langle \left[T_{xy}(t,{\bf q}),T_{xy}(0,-{\bf q})\right]\right\rangle \right\} ,\label{eq:eta}
\end{equation}
where $T_{xy}(t,{\bf q})$ is the off-diagonal part of the total stress tensor in momentum
space in the Heisenberg picture with respect to the two-channel Hamiltonian Eqs.~(\ref{eq:H0},
\ref{eq:Hint}).

In order to calculate the viscosity and avoid potential issues associated with analytic
continuation from imaginary axis \cite{kagamihara2019shear} we use the Keldysh formalism. The relevant diagrams, obtained by a formal perturbative expansion of the Kubo formula Eq.~\eqref{eq:eta} up to the $g^2$ order are shown in Fig.~\ref{Fig1}. As discussed in the Supplementary Material~\cite{Sup}, the
bosonic and fermionic Green's functions obey Dyson's equations with
the bosonic self-energy $\Pi$ and fermionic self-energy $\Sigma$ diagrammatically shown in Fig.~\ref{Fig1} (a).
We note that both self-energies are restricted to the lowest order
in perturbation theory in the coupling strength $g$. These self-energy
terms renormalize the mass and generate lifetime of quasiparticles.  
In the weak coupling limit, the latter provides the leading contribution to the viscosity. More specifically,
the lowest-order contributions to the viscosity that we denote as
$\eta_{F}$ and $\eta_{B}$, shown in Fig.~\ref{Fig1} (b),
can be expressed as follows:
\begin{align}
\eta_{F} & =\frac{1}{2Tm^{2}}\int\frac{d^{3}{\bf k}}{(2\pi)^{3}}(k_{x}k_{y})^{2}\text{sech}^{2}\left( \frac{\xi_{{\bf k}}^{F}}{2T} \right)\tau_{F}({\bf k},\xi_{{\bf k}}^{F}),\label{eq:etaa}\\
\eta_{B} & =\frac{1}{16Tm^{2}}\int\frac{d^{3}{\bf q}}{(2\pi)^{3}}(q_{x}q_{y})^{2}\text{csch}^{2}\left(\frac{\xi_{{\bf q}}^{B}}{2T}\right)\tau_{B}({\bf q},\xi_{{\bf q}}^{B}),\label{eq:etab}
\end{align}
where we defined the quasiparticle lifetimes through the retarded
components of the self-energies as follows $\tau^{-1}_{F}({\bf k},\epsilon)=-2\text{Im}\Sigma^{R}({\bf k},\epsilon)$,
$\tau^{-1}_{B}({\bf q},\omega)=-2\text{Im}\Pi^{R}({\bf q},\omega)$. Finite
viscosity therefore requires the on-shell lifetime of all quasiparticle
states to be finite. Due to kinematic constraints, the approximate self-energies (Fig.~\ref{Fig1}(a))
that we employ throughout this work, provide a finite lifetime only
for positive values of the detuning $\epsilon_{0}>0$.   

\begin{figure}[t!]
\begin{centering}
\includegraphics[width=1\columnwidth]{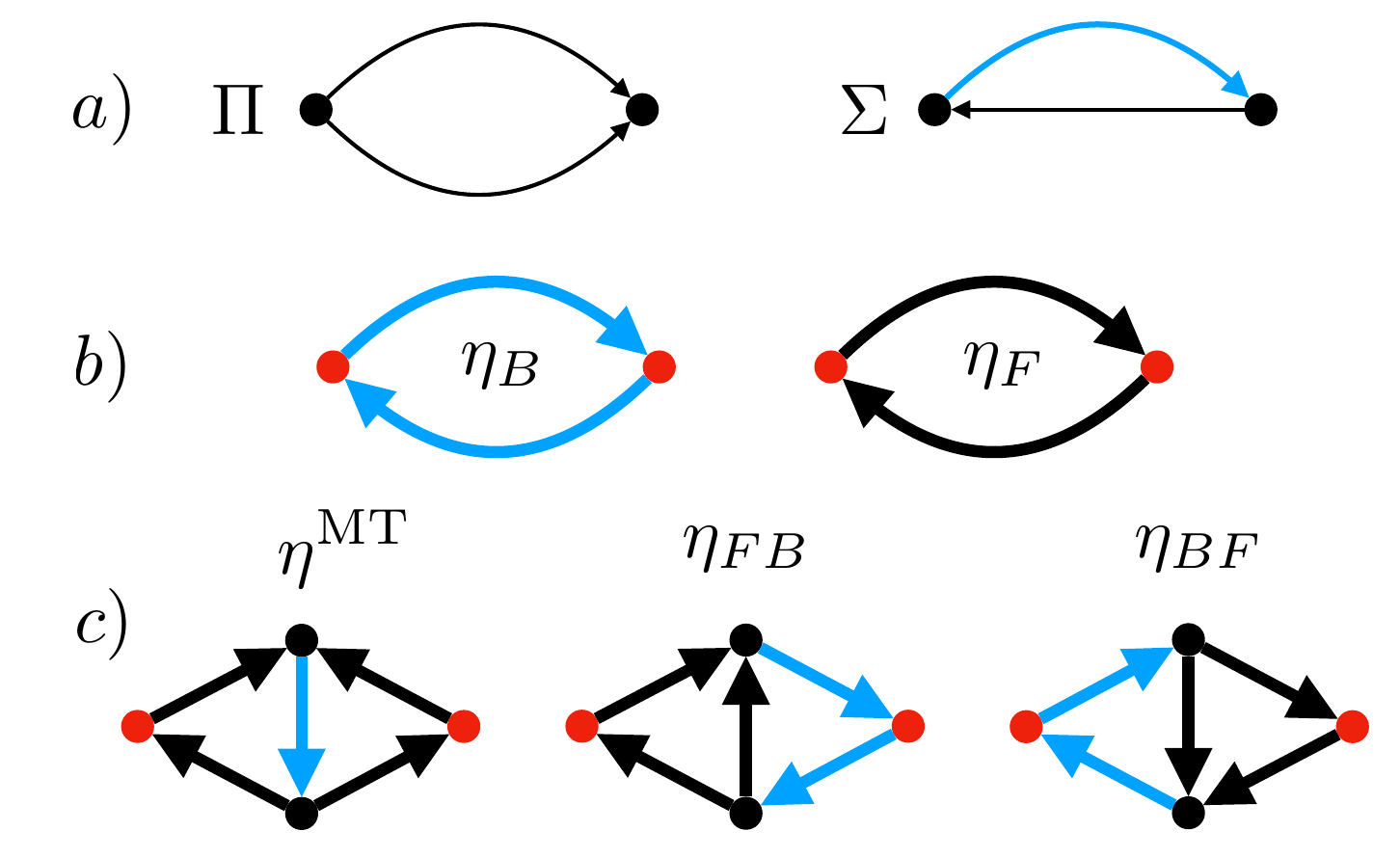} 
\par\end{centering}
\caption{Diagrammatic representation of perturbation theory for the linear
response in the two-channel model. Bosonic and fermionic propagators
are shown with blue and black arrows, respectively. Red circles denote
the viscosity vertices corresponding to Eqs.~\eqref{eq:TF}, \eqref{eq:TB} ($\sim k_x k_y$).
Black circles represent the coupling vertex in Eq.~\ref{eq:Hint}
and are proportional to $g$. (a) Relevant `self-energies'
for bosons ($\Pi$) and fermions ($\Sigma$). (b) Zeroth-order terms, obtained
by formal perturbative expansion of the Kubo formula Eq.~\eqref{eq:eta} in coupling strength $g$.
(c) Second-order $\sim g^2$  corrections, including Maki-Thompson ($\eta^{MT}$) and cross-species terms $\eta_{FB/BF}$. Thin and thick arrows represent bare and renormalized propagators respectively.}

\label{Fig1} 
\end{figure}
The lowest-order Drude-like terms, Eqs.~(\ref{eq:etaa}) and (\ref{eq:etab}),
are generally not sufficient to determine the viscosity. In fact,
while formally being higher-order in $g$, the higher-order corrections,
 required to satisfy the Ward identity associated with the momentum conservation law~\cite{baym1961conservation, enss2011viscosity}, are of the same order in $g$ as Eqs.~(\ref{eq:etaa}) and (\ref{eq:etab}). Examples
of such diagrams are shown in Fig.~\ref{Fig1} (c). In particular, $\eta^{\text{MT}}$ denotes the so-called Maki-Thompson contribution,
analogous to the theory of fluctuation corrections to conductivity
\cite{maki1989fluctuation,Fluct}. It generally describes coherent scattering of quasiparticles off of superconducting fluctuations.  The remaining terms, $\eta_{BF},\eta_{FB}$, effectively describe
the inter-species components of the viscosity. They stand for the
cross-response of the Bose-Fermi mixture, corresponding to the physical scenario
where the external perturbation is applied to the bosonic sector,
while the resulting momentum current is measured in the fermionic
sector (and vice versa). We note that we find numerically and demonstrate
analytically in the Supplementary Material~\cite{Sup} that $\eta_{FB}\approx\eta_{BF}$ in the limit of weak coupling $g$.
Diagrammatically, each of the terms $\eta_{FB},\eta_{BF}$ and $\eta^{\text{MT}}$
has one extra loop compared to the Drude-like terms in Eqs.~(\ref{eq:etaa})
and (\ref{eq:etab}). As we show in the SM, each such a loop generates
terms proportional to $\tau_{F/B}\sim g^{-2}$ that cancel the overall $g^{2}$
prefactor, thus resulting in the same order of magnitude contribution as Eqs.~(\ref{eq:etaa})
and (\ref{eq:etab}). 

The emergence of formally higher-order in $g$
  terms that are numerically comparable to lower-order contributions necessitates a self-consistent treatment of the Bethe-Salpeter equation for the viscosity vertex.
Standard approach to solving such equations is to employ the  imaginary time (Matsubara) formalism. This, however, entails the necessity of analytically continuing the result to real frequencies using
 Pad\'e approximants \cite{enss2011viscosity}. However at weak coupling such procedure can be unstable and is known to produce unphysical
results \cite{kagamihara2019shear}. We note that the same issue arises in the calculation
of conductivity in metals, and different approximate methods have
been developed to address it \cite{gotze1972homogeneous}. However, the regime of validity
of such approximations cannot be controlled.

\begin{figure}[t!]
\begin{centering}
\includegraphics[width=1\columnwidth]{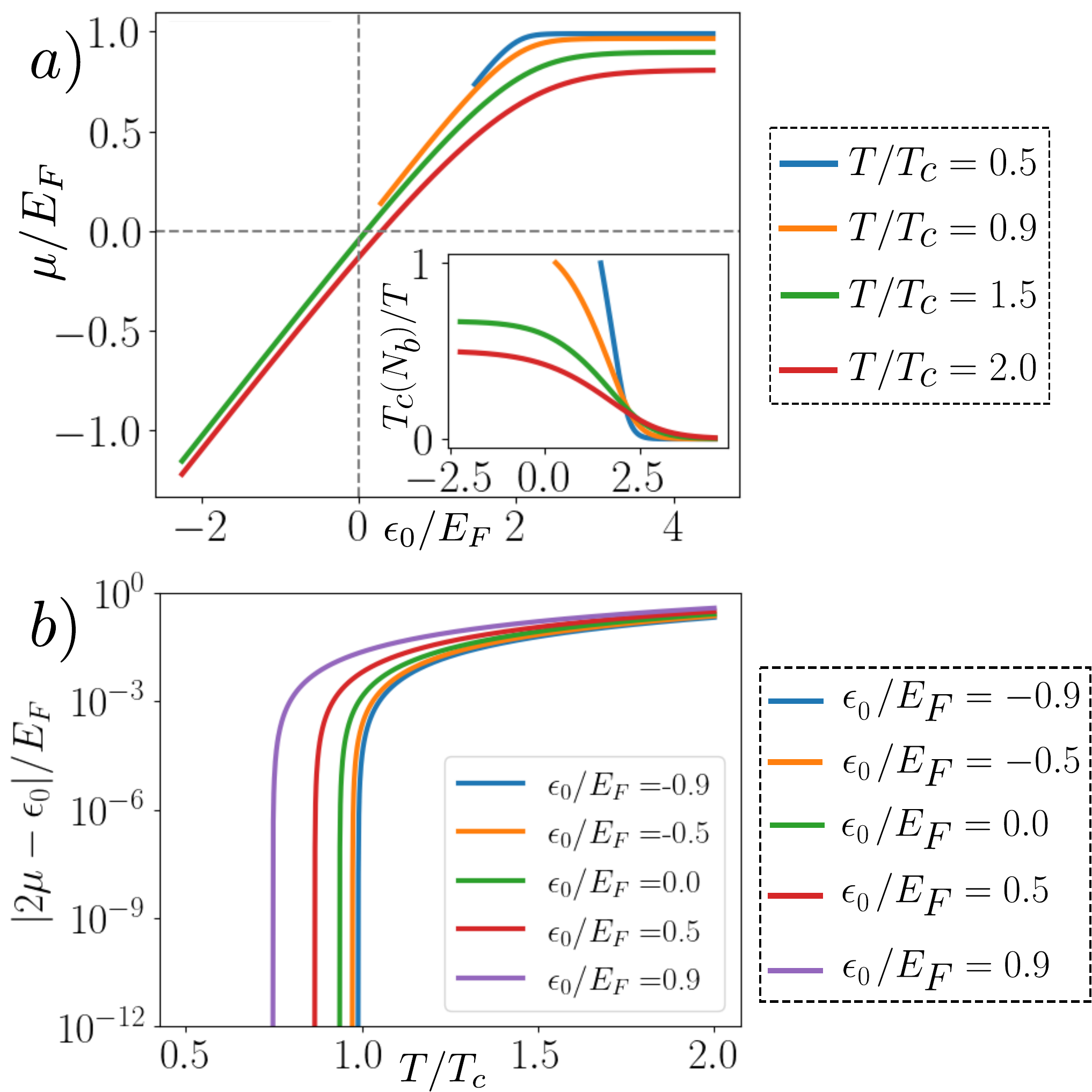} 
\par\end{centering}
\caption{Phase diagram of the two-channel model Eqs.~(\ref{eq:H0}, \ref{eq:Hint})
as function of  detuning $\epsilon_{0}$ and temperature $T$. (a) Fermionic chemical
potential at different temperatures expressed in units of BEC transition
temperature corresponding to the atom density. 
As the detuning varies, the system crosses over from the BCS regime at large positive detuning $\epsilon_0$, where $\mu\sim E_F$ and the system is dominated by free atoms, to the BEC regime at negative $\epsilon_0$, where $\mu<0$ and the system is dominated by bound molecules.
The inset shows that
the temperature $T_{c}(N_{b})\equiv\left(\frac{N_{b}}{\zeta(3/2)}\right)^{2/3}\frac{\pi}{m}$
corresponding to the bosonic density $N_{b}$, as a function of detuning. (b) Effective bosonic chemical potential $2\mu-\epsilon_{0}$
as function of the temperature for different values of detuning. In the $g\rightarrow 0$ limit, the critical temperature is determined by the condition this effective bosonic chemical potential vanishes.
}
\label{Fig1-1} 
\end{figure}

 In order to avoid the issues above, in this work we follow a different strategy. First, we employ the Keldysh formalism thereby avoiding the procedure of analytic continuation. 
Within this formalism, we compute the terms diagrammatically shown in Fig.~\ref{Fig1}.  We then restrict the validity of our theory to the regime where the
corrections obtained by the expansion up to the order $g^{2}$ ($\ensuremath{\eta_{BF},\eta_{FB}},\ensuremath{\eta^{\text{MT}}}$), computed with renormalized Green's functions,
are smaller than the lowest-order terms $\eta_{F/B}$
Before proceeding to the result of the calculation, we first explore
the phase diagram of the two-channel model Eqs.~(\ref{eq:H0}, \ref{eq:Hint}).
Within mean-field theory, the superfluid transition is generally controlled
by the particle number conservation and by a BCS-like equation encoding
the Thouless criterion \citep{thouless1960perturbation,nozieres1985} for condensation
$\epsilon_{0}-2\mu=g^{2}/2\int\frac{d^{3}{\bf k}}{(2\pi)^{3}}\tanh(\xi_{{\bf k}}^{F}/2T)/\xi_{{\bf k}}^{F}$. While the integral on the right-hand side of this equation is formally divergent without an ultraviolet cutoff, the divergent term can be absorbed \cite{gurarie2007resonantly} in the detuning. Without loss of generality, we assume the detuning is renormalized throughout this work. 
In the limit $g \rightarrow 0$, the BCS transition temperature becomes exponentially small and the transition temperature is predominantly controlled by 
the total atom conservation equation $N_{F}+2N_{B}=\text{const}$.
Furthermore, in the weak-coupling limit, we can ignore the fluctuation
corrections~\cite{Sup} to $N_{F}$ and $N_{B}$ and approximate the particle
distribution functions with the bare non-interacting Bose and Fermi
distribution functions, i.e. $N_{F}=2\int \frac{d^{3}{\bf k}}{(2\pi)^2} n_{F}({\bf k})$ and $N_{B}=\int \frac{d^{3}{\bf k}}{(2\pi)^2} n_{B}({\bf k})$,
where $n_{F/B}({\bf k})=(\exp\{\beta\xi_{{\bf k}}^{F/B}\}\pm1)^{-1}$.
Under these approximations, the transition temperature is determined
when the equation $N_{F}+2N_{B}=\text{const}$ does not have a solution
as a function of $\mu$. We note that $\mu$, being the fermionic
chemical potential, can take both positive and negative values. The
effective bosonic chemical potential $2\mu-\epsilon_{0}$, however,
can only take negative values. Taking these considerations into account,
the phase diagram is fully determined by the temperature and the detuning
$\epsilon_{0}$ that can be changed by varying the external magnetic
field.

\begin{figure}[]
\begin{centering}
\includegraphics[width=1\columnwidth]{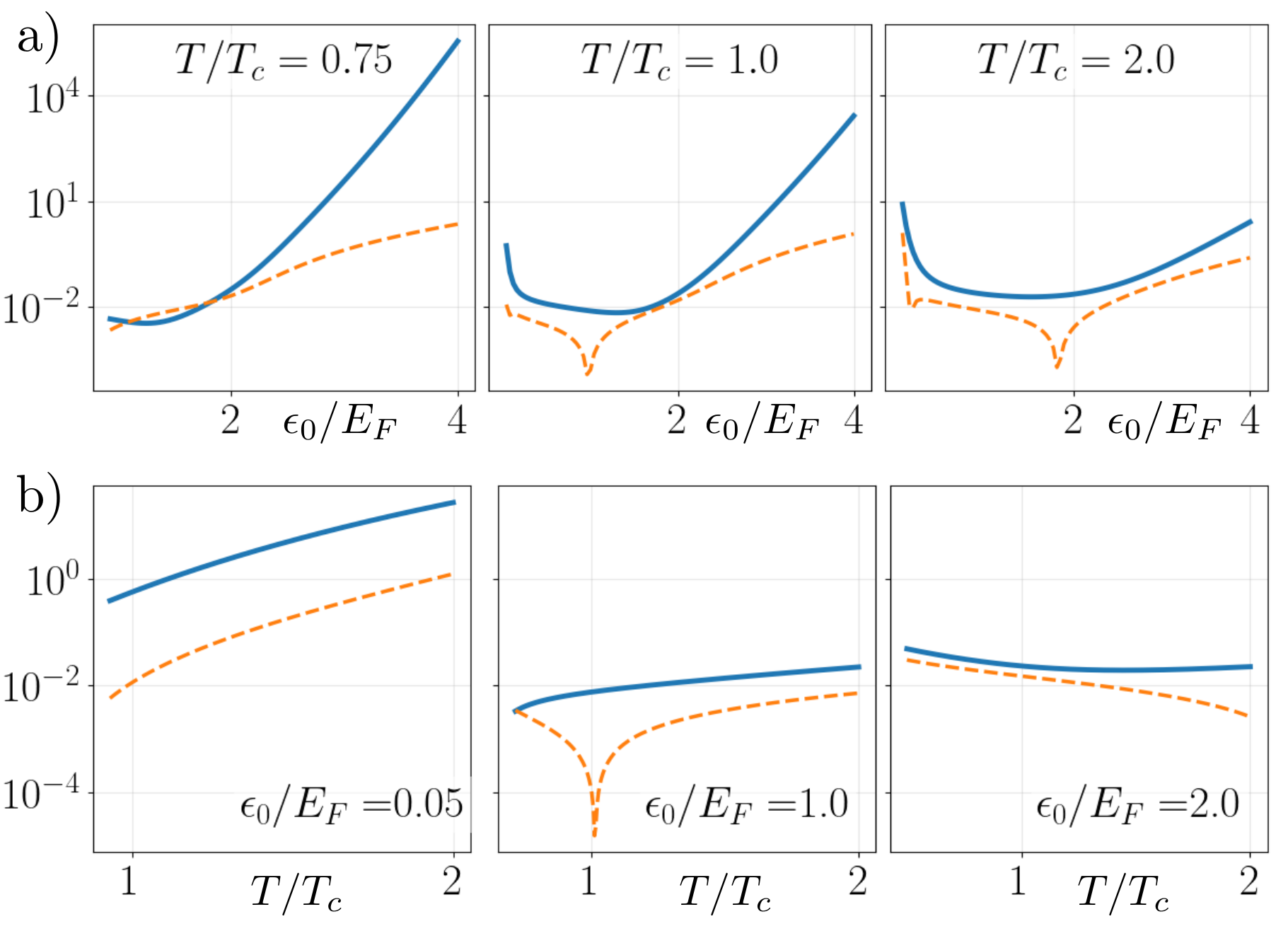} 
\par\end{centering}
\caption{Shear viscosity $\eta$ in the two-channel model Eqs.~(\ref{eq:H0}, \ref{eq:Hint}).
(a) $\eta$ as function of detuning $\epsilon_{0}$ at different temperatures.
(b) $\eta$ as function of temperature at different detunings $\epsilon_{0}$.
The blue curve corresponds to the total viscosity $\eta_{\rm total}$ defined in Eq.\eqref{eq:eta-total},
while the orange dashed curve is the absolute value of the lowest-order correction 
$\eta^{MT}+\eta_{FB}+\eta_{BF}$, which, when added to the naive estimation $\eta_F+\eta_B$, gives the total viscosity. The cusp-like features are an artifact of the logarithmic scale in the y-axis, corresponding to near-zero corrections.
The range of the shown detuning/temperature values corresponds to the phase diagram Fig.~\ref{Fig1-1} (a, b) and is constrained
by the condensation of bosons (see Fig.~\ref{Fig0}~(a)) and the
$\epsilon_{0}>0$ condition. }

\label{Fig3} 
\end{figure}
We now explore the phase diagram fixing the fermionic density $n$
and temperature $T$. It is convenient to define the Fermi energy
if all atoms were fermionic as $E_{F}=\left(3n\pi^{2}\right)^{2/3}/2m$,
and the BEC transition temperature if all atoms were converted to
bosons as $T_{c}=\left(\frac{n}{2\zeta(3/2)}\right)^{2/3}\frac{\pi}{m},$
where $\zeta$ is the Riemann Zeta function. As can be seen in Fig.~\ref{Fig1-1}~(a)
at temperatures  $T>T_{c}$,
as a function of $\epsilon_{0}$,
the system can crossover from the BEC regime containing mostly bosonic molecules ($\mu<0$) to the BCS regime containing mostly fermionic atoms ($\mu\sim E_F$).
At $T<T_{c}$ the system undergoes a Bose-Einstein condensation
when the density of bosons $N_{b}$ matches the critical one as
shown on the inset of Fig.~\ref{Fig1-1}~(a). We note that generally
the transition temperature, which in the $g \rightarrow 0$ limit is determined by the condition that effective bosonic potential vanishes, $2\mu-\epsilon_0=0$, decreases for positive values
of $\epsilon_{0}$ as can be seen in Fig.~\ref{Fig1-1}~(b).

 We now numerically evaluate the five perturbative contributions to the shear viscosity, illustrated via the diagrams in Fig.~\ref{Fig1}~(b, c). The results are summarized in Fig.~\ref{Fig3}~(a, b): panel (a) shows $\eta$ as a function of the detuning $\epsilon_{0}$ for various temperatures above and below the BEC transition ($T/T_{c}$), while panel (b) shows $\eta$ as a function of temperature for different detunings.
 We find that the minimum of the total viscosity 
\begin{equation}\label{eq:eta-total}
    \eta_{\rm total} = \eta_F + \eta_B + \ensuremath{\eta^{\text{MT}}} + \ensuremath{\eta_{BF} + \eta_{FB}},
\end{equation}
is generally located away from the resonance at $\epsilon_{0}=0$ \cite{kagamihara2019shear}. We note, however, that the near-resonant regime is beyond the range of applicability of perturbation theory. In particular, we find that in this case, the higher-order corrections to $\eta$ are generally of the same order of magnitude as the Drude-like contributions, $\eta_{F}+\eta_{B}$. In this regime, this implies that a full self-consistent treatment of the vertex corrections may be necessary to provide a definitive quantitative description. We also observe that in this non-perturbative regime $\epsilon_0\approx 0$, our approximate self-energies produce divergent viscosity due to the fact that the quasiparticle lifetime is infinite. At larger values of detuning, the viscosity increases due to the decreasing effective interaction strength; in this regime, the Drude-like terms [Eqs.~\eqref{eq:etaa} and \eqref{eq:etab}] provide the dominant contribution. Crucially, our results demonstrate that $\eta$ is widely tunable via $\epsilon_0$, providing an experimental knob to explore the transition to turbulence in ultracold atomic systems.

In conclusion, we have characterized the shear viscosity of an ultracold Fermi gas across the BCS-BEC crossover. Using a two-channel model in the weak-coupling limit, we demonstrate that the viscosity is highly sensitive to the magnetic field, allowing it to be tuned over several orders of magnitude. An interesting future research direction would involve moving beyond the weak-coupling perturbative expansion by employing the full numerical solution of Kadanoff-Baym equations on Keldysh contour. In particular, this approach would enable a direct comparison with  $\phi^4$  results in the deep BEC limit~\citep{phi4,phi4-1,phi4-2}. 
Furthermore, this would allow for an extension of our results beyond linear response theory and could provide novel insights into the dynamics of cloud expansion experiments \citep{cao2011universal}.
\paragraph{Acknowledgements.} All authors are grateful to Martin Zwierlein and his group for numerous discussions and collaboration related to the experimental realization of the tunable viscosity. This work was supported by DARPA APAQuS (AG, AP, VG, LL). YL acknowledges support from the Knut and Alice Wallenberg Foundation (KAW 2024.0131). We thank Victor Gurarie and Leo Radzihovsky for discussions.



\bibliography{bibl}

\end{document}


\title{Tunable viscosity across the BCS-BEC crossover
    \\
    Supplemental Material
    }

\author{Yunxiang Liao}
\affiliation{Department of Physics, KTH Royal Institute of Technology, SE-106 91
Stockholm, Sweden}

\author{Andrey Grankin}
\affiliation{Joint Quantum Institute, Department of Physics, University of Maryland,
College Park, MD 20742, USA}

\author{Archisman Panigrahi}
\affiliation{Massachusetts Institute of Technology, Cambridge, MA USA}

\author{Victor Galitski}
\affiliation{Joint Quantum Institute, Department of Physics, University of Maryland,
College Park, MD 20742, USA}

\author{Leonid Levitov}
\affiliation{Massachusetts Institute of Technology, Cambridge, MA USA}

	\date{\today}

	\maketitle


 \section{Two-channel model}

In this supplementary material, we provide the detailed derivation of the shear viscosity of a Fermi gas with Feshbach resonance.
We focus on the narrow resonance limit, and consider the s-wave two channel model  described by the Hamiltonian
\begin{align}
\begin{aligned}
    H=\,&
    \sum_{\sigma,\pb}
    \left( \frac{p^2}{2m}-\mu \right) 
    \psi^{\dagger}_{\sigma}  (\pb)
    \psi_{\sigma}  (\pb)
    +
    \sum_{\qb}
    (\varepsilon_0+ \frac{q^2}{4m}-2\mu )\phi^{\dagger} (\qb) \phi (\qb)
     \\
    &+
    g
    \sum_{\qb,\pb}
    \left(
    \phi^{\dagger} (\qb)
    \psi_{\downarrow}  (-\pb+\frac{\qb}{2})
    \psi_{\uparrow}  (\pb+\frac{\qb}{2})
    +
    \phi (\qb)
    \psi_{\uparrow}^{\dagger}  (\pb+\frac{\qb}{2})
    \psi_{\downarrow}^{\dagger}  
    (-\pb+\frac{\qb}{2})
    \right).
\end{aligned}
\end{align}
where $\psi_{\sigma}$ denotes the annihilation operator for open-channel fermionic atoms with pseudospin $\sigma=\uparrow,\downarrow$, while $\phi$ represents that of the close-channel bosonic molecule. $m$ is the atomic mass, $\mu$ is the fermionic chemical potential and $\e_0$ is the detuning.
The coupling constant $g$ determines the width of Feshbach resonance, and is small for the narrow resonance. Therefore, it serves as a small parameter in our perturbative calculation.

From the Kubo formula, we know that shear viscosity $\eta$ can be obtained from the retarded correlation function $\GG^{(R)}_{T}(\omega,\qb)$ of the stress tensor $T_{xy}$:
\begin{align}\label{eq:Kubo}
						\begin{aligned}
							&\eta
							=-\,
							\lim\limits_{\omega \rightarrow 0}
							\lim\limits_{\qb \rightarrow \vex{0}}
							\frac{1}{\omega}
							\operatorname{Im}
							\GG^{(R)}_{T}(\omega,\qb),	 
							\\
							&\GG^{(R)}_{T}(\omega,\qb)
							=\,
							-i
							\int_{-\infty}^{+\infty}dt
							\int d\rb
							\,
							e^{i\omega t - i \qb \cdot \rb}
							\Theta(t)
							\braket{
								\left[ 
								T_{xy}(\rb,t),
								T_{xy}(\vex{0},0)
								\right] 
							}.
\end{aligned}
\end{align}

For the two-channel model, 
the stress tensor $T_{ij}=T^{(F)}_{ij}(\rb,t)+T^{(B)}_{ij}(\rb,t)+T^{(int)}_{ij}(\rb,t)$ consists of three components:
						\begin{subequations}
                        \begin{align}
							&\begin{aligned}\label{eq:TF}
								& T^{(F)}_{ij}(\rb,t) =
								\frac{1}{2m}
								\sum_{\sigma}
								\left[ 
								\partial_i \psi^{\dagger}_{\sigma} (\rb,t)
								\partial_j \psi_{\sigma} (\rb,t)
								+
								\partial_j \psi^{\dagger}_{\sigma} (\rb,t)
								\partial_i \psi_{\sigma} (\rb,t)
								-\frac{1}{2}
								\delta_{ij} \nabla^2 \left( \psi^{\dagger}_{\sigma}(\rb,t)  \psi_{\sigma}(\rb,t) \right) 
								\right],
                                \end{aligned}	
								\\
                               & \begin{aligned}\label{eq:TB}
								&T^{(B)}_{ij}(\rb,t)
								=\,
								\frac{1}{4m}
								\left[ 
								\partial_i \phi^{\dagger} (\rb,t)
								\partial_j \phi (\rb,t)
								+
								\partial_j \phi^{\dagger} (\rb,t)
								\partial_i \phi (\rb,t)
								-\frac{1}{2}
								\delta_{ij} \nabla^2 \left( \phi^{\dagger}(\rb,t)  
								\phi(\rb,t) \right) 
								\right],
                                \end{aligned}	
                                \\
                               & \begin{aligned}\label{eq:Tint}
								&T_{ij}^{(\msf{int})}(\rb,t)
								=
								\frac{g}{2}	
								\delta_{ij}
								\left[
								\phi^{\dagger}(\rb,t)
								\psi_{\downarrow}  (\rb,t) 
								\psi_{\uparrow} (\rb,t)
								+
								\phi(\rb,t)
								\psi^{\dagger}_{\uparrow}  (\rb,t)
								\psi_{\downarrow}^{\dagger}  (\rb,t)
								\right].
                                \end{aligned}	
							\end{align}	
						\end{subequations}
Here the subscript $i,j=x,y,z$ are the Cartesian indices. $T^{(F)}_{ij}$ and $T^{(B)}_{ij}$ are the stress tensors for noninteracting fermions and bosons with dispersion $\xi_k^{(F)}=k^2/2m-\mu$ and $\xi_q^{(B)}=q^2/4m-2\mu+\e_0$, respectively, while $T^{(\msf{int})}_{ij}$ represents the additional contribution from boson-fermion interactions. We emphasize that the shear viscosity, determined by the off-diagonal stress tensor component $T_{xy}$ in the Kubo formula, receives contributions only from the noninteracting stress tensors $T_{xy}^{(F/B)}$.

 \section{Stress tensor derivation}

The derivation of the stress tensor for the two channel model is as following.
Since two fermionic atoms can bind into a bosonic molecule, and one bosonic molecule can convert into two fermionic atoms, the boson and fermion momenta are not separately conserved. However, their summation - the total momentum - is conserved and satisfies the momentum conservation equation
\begin{align}\label{eq:momentum}
                        &\frac{\partial g_i}{\partial t} 
						+
						\partial_j T_{ji}
						=0.
\end{align}
Here $g_i$ is the total momentum density in Cartesian direction $i$, and is composed of fermionic part $g_i^{(F)}$ and bosonic part $g_i^{(B)}$:
\begin{align}
\begin{aligned}
    g_i^{(F)}(\rb,t)
						=\,&
						\frac{1}{2i}
                        \sum_{\sigma}
						\left[ 
						\psi^{\dagger}_{\sigma} (\rb,t)
						\partial_i
						\psi_{\sigma} (\rb,t)
						-
						\left( 
						\partial_i
						\psi^{\dagger}_{\sigma} (\rb,t)
						\right) 
						\psi_{\sigma} (\rb,t)
						\right],
                        \\
    g_i^{(B)}(\rb,t)
						=\,&
						\frac{1}{2i}
						\left[ 
						\phi^{\dagger} (\rb,t)
						\partial_i
						\phi (\rb,t)
						-
						\left( 
						\partial_i
						\phi^{\dagger} (\rb,t)
						\right) 
						\phi (\rb,t)
						\right].                  
\end{aligned}
\end{align}
We now make use of the equation of motion,
					\begin{align}
						\begin{aligned} 
							i \frac{\partial}{\partial t} \psi_{\sigma} (\rb, t)
							=\,&
							-[H, \psi_{\sigma} (\rb, t)]
							=
							-\frac{1}{2m}
							\nabla^2 \psi_{\sigma}  (\rb,t)
                            -\mu \psi_{\sigma}  (\rb,t)
							+
							gs_{\sigma}
							\phi(\rb,t)\psi_{\bar{\sigma}}^{\dagger}  (\rb,t),
                            \\
                            i \frac{\partial}{\partial t} \phi (\rb, t)
							=\,&
							-[H, \phi (\rb, t)]
							=
							-\frac{1}{4m}
							\nabla^2 \phi  (\rb,t)
							+
							(\e_0-2\mu)\phi  (\rb,t)
							+
							\phi(\rb,t)
							\psi_{\downarrow} (\rb,t)
							\psi_{\uparrow}  (\rb,t),
						\end{aligned}   
					\end{align}
where $s_{\uparrow}=-s_{\downarrow}=1$, and $\bar{\sigma}$ denotes the spin opposite to $\sigma$. 
This leads to
						\begin{align}
							\begin{aligned}
								&\frac{\partial }{\partial t}g_i^{(F/B)}(\rb,t)
								+
								\partial_j T^{(F/B)}_{ji}(\rb,t)
								=
								F^{(F/B)}_i,
							\end{aligned}	
						\end{align}               
                        where $T^{(F/B)}$ is the noninteracting fermionic/bosonic stress tensor, whose explicit expressions is given by Eq.~\ref{eq:TF}/\ref{eq:TB},
                        and $F$ represents momentum transfer from boson-fermion coupling:
									\begin{align}
							\begin{aligned}
								F^{(F)}_i
								=&
								-
								g
								\left( 
								\partial_{i}
								\phi^{\dagger}(\rb,t)
								\right)
								\psi_{\downarrow}  (\rb,t) 
								\psi_{\uparrow} (\rb,t)
								-  
								g
								\left( 
								\partial_{i} 
								\phi(\rb,t)
								\right)
								\psi^{\dagger}_{\uparrow}  (\rb,t)
								\psi_{\downarrow}^{\dagger}  (\rb,t)
                                ,
								\\
								F^{(B)}_i
								=&
								+
								\frac{g}{2}
								\left(   
								\partial_{i} \phi (\rb,t)
								\right) 
								\left(   
								\psi_{\uparrow}^{\dagger}  (\rb,t)
								\psi_{\downarrow}^{\dagger} (\rb,t)
								\right) 
								-
								\frac{g}{2}
								\phi (\rb,t)
								\partial_{i}
								\left(     
								\psi_{\uparrow}^{\dagger}  (\rb,t)
								\psi_{\downarrow}^{\dagger} (\rb,t)
								\right) 
								\\
								&-
								\frac{g}{2}
								\phi^{\dagger}  (\rb,t)
								\partial_{i}
								\left( 
								\psi_{\downarrow} (\rb,t)
								\psi_{\uparrow}  (\rb,t)
								\right) 
								+
								\frac{g}{2}
								\left( 
								\partial_{i}    
								\phi^{\dagger} (\rb,t)
								\right) 
								\left( 
								\psi_{\downarrow} (\rb,t)
								\psi_{\uparrow}  (\rb,t)
								\right).
							\end{aligned}	
						\end{align}  
From the total momentum conservation equation Eq.~\ref{eq:momentum}, we find that the interaction contribution to the stress tensor $T_{ij}^{(\msf{int})}$ obeys
\begin{align}
                -\partial_j T_{ji}^{(\msf{int})}(\rb,t)
								=
								F^{(F)}_i+F^{(B)}_i
                                =
                                -
								\frac{g}{2}	
								\partial_{i}
								\left(
								\phi^{\dagger}(\rb,t)
								\psi_{\downarrow}  (\rb,t) 
								\psi_{\uparrow} (\rb,t)
								+
								\phi(\rb,t)
								\psi^{\dagger}_{\uparrow}  (\rb,t)
								\psi_{\downarrow}^{\dagger}  (\rb,t)
								\right),
\end{align}
which leads to Eq.~\ref{eq:Tint}.

\section{Keldysh formalism for the two channel model}

In this section, we set up the Keldysh formalism 
for the two channel model, 
					\begin{align}
						\begin{aligned}
							\label{eq:Z1}
							&Z
							\equiv
							\int 
							\D \left( \bar{\psi}, \psi \right) 
							\D \left( \bar{\phi}, \phi \right) 
							\,
							\exp
							\left( 
							iS_\psi+iS_\phi+iS_{\msf{int}}
							\right) ,
							\\
							&i S_\psi
							=
							i 
							\sum_{\sigma=\uparrow,\downarrow}
							\int_{\rb,t}
							\bar{\psi}_{\sigma}(\rb,t)
							\left(i\partial_t + \frac{\nabla^2}{2m} + \mu \right)
							\psi_{\sigma}(\rb,t),
							\\
							&i S_\phi
							=
							i \int_{\rb,,t}
							\bar{\phi}(\rb,t)
							\left(i\partial_t + \frac{\nabla^2}{4m} +2\mu -\varepsilon_0 \right)
							\phi(\rb,t),
							\\
							&iS_{\msf{int}}
							=
							-ig
							\sum_{a=+,-}
							\tau^3_{aa}
							\int_{\rb}
							\left[
							\bar{\phi}^a(\rb,t)
							\psi^a_{\downarrow}(\rb,t)
							\psi^a_{\uparrow}(\rb,t)
							+
							\phi^a(\rb,t)
							\bar{\psi}^a_{\uparrow}(\rb,t)
							\bar{\psi}^a_{\downarrow}(\rb,t)
							\right].
						\end{aligned}
					\end{align}
					Here
					$ \psi_{\sigma}^{a}$ represents the fermionic atom and carries Keldysh 
					$a \in \{+,-\}$ and spin $\sigma \in \{\uparrow,\downarrow\}$ labels.
					$\phi^{a}$ represents the bosonic molecule  and carries only a Keldysh index $a$.
					$\hat{\tau}$ stands for the Pauli matrix acting on the Keldysh space.

As mentioned, the shear viscosity can be obtained from the correlation function of the stress tensor $T_{xy}$. We define the classical and quantum components of the stress tensor $T_{xy}$:
\begin{align}
					T_{xy}^{cl}=\left( T_{xy}^{+}+T_{xy}^{-}\right) /\sqrt{2},
					\qquad
					T_{xy}^{q}=\left( T_{xy}^{+}-T_{xy}^{-}\right) /\sqrt{2}.
\end{align}
Here $T_{xy}^{\pm}$ is the stress tensor in the forward/backward branch, and is composed of the fermionic and bosonic components
						\begin{align}
							\begin{aligned}
								& T^{(F)\pm}_{xy}(\rb,t) =
								\frac{1}{2m}
								\sum_{\sigma}
								\left[ 
								\partial_x \bpsi_{\sigma}^{\pm} (\rb,t)
								\partial_y \psi_{\sigma}^{\pm} (\rb,t)
								+
								\partial_y \bpsi_{\sigma}^{\pm} (\rb,t)
								\partial_x \psi_{\sigma}^{\pm} (\rb,t)
								\right],
								\\
								&T^{(B)\pm}_{xy}(\rb,t)
								=\,
								\frac{1}{4m}
								\left[ 
								\partial_x \bphi^{\pm} (\rb,t)
								\partial_y \phi^{\pm} (\rb,t)
								+
								\partial_y \bphi^{\pm} (\rb,t)
								\partial_x \phi^{\pm} (\rb,t)
								\right].
							\end{aligned}	
						\end{align}
Note that the interaction contribution  to the stress tensor  disappears for the $xy$ component $T^{(\msf{int})}_{xy}=0$.             
The retarded correlation function of the stress tensor is given by
					\begin{align}\label{eq:GT-0}
						\begin{aligned}
							&
							&\GG^{(R)}_{T}(\Qb,\WW)
							=\,
							-i
							\braket{
								T_{xy}^{cl}(\Qb,\WW)
								T_{xy}^{q}(-\Qb,-\WW)
							},
						\end{aligned}
					\end{align}
from which one can obtain the shear viscosity by taking the zero external momentum and zero external frequency limits (Eq.~\ref{eq:Kubo}).

We introduce the Nambu spinor $\Psi^{\pm} = \begin{bmatrix}
                        \psi_{\uparrow}^{\pm}
                        &
                        \bpsi_{\downarrow}^{\pm}
                    \end{bmatrix}$,
                    and then apply the Keldysh rotation for the Nambu spinor
                    \begin{align}\label{CoV}
						\begin{aligned}
							&
                            \Psi
                            =
                            \begin{bmatrix}
                            \Phi^1
                            &
                            \Phi^2
                            \end{bmatrix}^T,
                            \qquad 
							\Psi^{1/2}
							\equiv
							\left( \Psi^+ \pm \Psi^-\right) /\sqrt{2},
					        \qquad
							\bar{\Psi}^{1/2}
							\equiv
							\left( \bar{\Psi}^+ \mp \bar{\Psi}^-\right) /\sqrt{2}.
						\end{aligned}
					\end{align} 
and also for the bosons
                    	\begin{align}
						\begin{aligned}
							&
                            \phi
                            =\begin{bmatrix}
                            \phicl
                            &
                            \phiq
                            \end{bmatrix}^{T},
                            \qquad
                            \phi_{\msf{cl}/\msf{q}}
							\equiv
							\left( \phi_+ \pm \phi_-\right) /2,
							\qquad
							\bphi_{\msf{cl}/\msf{q}}
							\equiv
							\left( \bphi_+ \pm \bphi_-\right) /2.
						\end{aligned}
					\end{align}
This leads to the actions
                \begin{align}
                \begin{aligned}\label{eq:Z0}
    								&i S_\psi
    								=
    								i
    								\int_{\pb,\e}
                                   \bar{\Psi}(\pb,\e)
    								G_0^{-1}(\pb,\e)
                                    {\Psi}(\pb,\e),
    								\\
    								&i S_\phi
    								=
    								i \int_{\qb,\ww}
                                    \bar{\phi}(\qb,\ww)
                                    D_0^{-1}(\qb,\ww) 
                                    \phi(\qb,\ww),
    								\\
    								&iS_{\msf{int}}
    								=
    								-ig 
    								\int_{\pb,\e,\qb,\ww}
                                     \bar{\Psi}(\pb+\qb,\e+\ww) 
                                     \Phi(\qb,\ww)
                                     {\Psi}(\pb,\e). 
    								\end{aligned}
      						\end{align}    
Here $\Phi$ represents the matrix 
                                  \begin{align}\label{eq:Phi}
                                  \begin{aligned}
                                      \Phi=(\phicl  +\phiq \tau_1)\otimes \sigma^+
                                            +(\bphicl  + \bphiq \tau_1)\otimes\sigma^-,
                                  \end{aligned}
                                  \end{align}
with $\sigma^{\pm}=(\sigma^1\pm i \sigma^2)/2$ being the raising/lowering Pauli matrix in the Nambu space.
The fermionic Green's function $G_0$  acquires the standard causality structure in the Keldysh space,
\begin{align}\label{eq:causality}
							G_0
							=
							\begin{bmatrix}
								G_0^{(R)} & G_0^{(K)}
								\\
								0 & G_0^{(A)}
							\end{bmatrix}_{\tau},
\end{align}
with $G^{(K)}_0$ satisfying the fluctuation-dissipation relation
\begin{align}\label{eq:FDT-f}
    								G_0^{(K)}(\kb,\e)
    								=
    								\left(
    								G_0^{(R)}(\kb,\e)-G_0^{(A)}(\kb,\e)
    								\right)
    								\tanh \left(\e/2T \right).
\end{align}
$G_0$ is diagonal in the Nambu space, 
                        \begin{align}
						\begin{aligned}
							G_0
							=
							\begin{bmatrix}
								G_{0p} & 0
								\\
								0 & G_{0h}
							\end{bmatrix}_{\sigma}.
						\end{aligned}
					\end{align}
We have used the subscript $\tau$/$\sigma$ to indicate the Keldysh/Nambu space. $G_{0p}$ and $G_{0h}$ denote the particle and hole components of $G_0$, respectively. Their retarded/advanced components are given by

                                \begin{align}\label{eq:G0-1}
    							\begin{aligned}
    								& G_{0p}^{(R/A)}(\kb,\e)	=\left[\e - (\frac{k^2}{2m}-\mu) \pm i \eta  \right]^{-1},
    								\qquad
    									G_{0h}^{(R/A)}(\kb,\e)	=\left[\e + (\frac{k^2}{2m}-\mu) \pm i \eta  \right]^{-1},
							\end{aligned}
						\end{align} 
where $\eta\rightarrow 0^+$ is a positive infinitesimal. 
                        
The bosonic Green's function assumes the form
\begin{align}\label{eq:causality2}
\begin{aligned}
    	D_0
							=
							\begin{bmatrix}
								D_0^{(K)} & D_0^{(R)}
								\\
								D_0^{(A)} & 0
							\end{bmatrix},
\end{aligned}
\end{align}
where

                                \begin{align}
    							\begin{aligned}
    								&
    								D_0^{(R/A)}(\qb,\ww)	
    								=
    								\frac{1}{2}
    								\left[\ww - (\frac{q^2}{4m}-2\mu +\e_0)   \pm i \eta  \right]^{-1},
							\end{aligned}
						\end{align}    
                        and $D^{(K)}_0$ can be determined from the fluctuation-dissipation relation
    								\begin{align}\label{eq:FDT-b}
    							\begin{aligned}
    								D^{(K)}_0(\omega,\textbf{q})
    								=
    								\left(
    								D^{(R)}_0(\omega,\textbf{q})
    								-
    								D^{(A)}_0(\omega,\textbf{q})
    								\right)
    								\coth \left(\omega/2T\right).
							\end{aligned}
						\end{align}

After the Keldysh rotation, the stress tensors are given by
					\begin{align}
						\begin{aligned}\label{eq:Txy-Keldysh}
							T^{(F)cl/q}_{xy}(\vex{0},\WW)
							=\,&
							\frac{1}{\sqrt{2}}
							\int_{\kb,\e}
							\frac{k_x k_y}{m}
							\bar{\Psi} (\kb,\e)
							\hat{\gamma}_{cl/q}^{(F)} \sigma^3
							\Psi (\kb,\e+\WW),
							\\
							T^{(B)cl/q}_{xy}(\vex{0},\WW)
							=\,&
							\sqrt{2}
							\int_{\qb,\e}
							\frac{q_x q_y}{2m}
							\bar{\phi} (\qb,\ww)
                            \hat{\gamma}_{cl/q}^{(B)}
							\phi(\qb,\ww+\WW),
						\end{aligned}
					\end{align}
where  $\hat{\gamma}_{cl}^{(F)}=\hat{\gamma}_{q}^{(B)}=\hat{\tau}_1$, and $\hat{\gamma}_{q}^{(F)}=\hat{\gamma}_{cl}^{(B)}=\hat{1}$.

 \section{Nozi\`eres-Schmitt-Rink theory for the two-channel model}

Integrating out the fermions, we arrive at an effective action for the bosons
					\begin{align}
						\begin{aligned}
							\label{eq:Seff}
							&Z
							\equiv
							\int 
							\D \left( \bar{\phi}, \phi \right) 
                            e^{iS_{\msf{eff}}[\bar{\phi},\phi]},
							\\
							&S_{{\msf{eff}}}[\bar{\phi},\phi]
                            =
							\int_{\qb,\ww}
                                    \bar{\phi}(\qb,\ww)
                                    D_0^{-1}(\qb,\ww) 
                                    \phi(\qb,\ww)
                            +i
							\Tr \ln
							G_{\phi},
						\end{aligned}
					\end{align}
 where we have defined
 \begin{align}
 \begin{aligned}\label{eq:Gphi}
    \G_{\phi}
    =
    (G_0^{-1}-g\Phi)^{-1} .
 \end{aligned}
 \end{align}
Note that $\Phi$ is a matrix  in the Keldysh and Nambu space defined in Eq.~\ref{eq:Phi}.

 \subsection{Saddle point equation}
 
The saddle point equation $\delta S_{\msf{eff}}/\delta \bar{\phi}_{\msf{q}}=0$ evaluated at the spatially uniform ansatz $\phicl=\Delta$ and $\phiq=0$ leads to
						\begin{align}
						\begin{aligned}
							&i
							(D_0^{(R)})^{-1}(0,0) 
							=
                            g\Tr \left(G_{\phi}\lvert_{\phicl=\Delta,\phiq=0}(\tau_1\otimes \sigma^-)\right),
                            	\end{aligned}
					\end{align}	
which can be further simplified to 
\begin{align}\label{eq:SPEQ}
						\begin{aligned}
                            \frac{i}{2}
    							 (2\mu -\e_0)  
                            =g^2
							\int_{\kb,\e}
                            \mathcal{F}_p^{(K)}(\kb,\e).
						\end{aligned}
					\end{align}						
 Here $\mathcal{F}_{p/h}$ represents the anomalous components (off-diagonal components in the Nambu space) of the Green's function $G_{\phi}\lvert_{\phicl=\Delta,\phiq=0}$, whose normal (diagonal) component is denoted by $\mathcal{G}_p$ in the following:
 \begin{align}\label{eq:Gphi-2}
 \begin{aligned}
             &   G_{\phi}\lvert_{\phicl=\Delta,\phiq=0}
             =\begin{bmatrix}
								\hat{G}_{0p}^{-1} & -g\Delta 
								\\
								-g \bar{\Delta}  & \hat{G}_{0h}^{-1} 
							\end{bmatrix}^{-1}    
                            =
							\begin{bmatrix}
								\mathcal{G}_p
								&  g \Delta \mathcal{F}_p
								\\
								g \bar{\Delta}\mathcal{F}_h
								& 
								\mathcal{G}_h
							\end{bmatrix},
                            \\
                                    &\mathcal{F}_p\equiv
                                    \left[ \hat{G}_{0h}^{-1}(\kb,\e)\hat{G}_{0p}^{-1} (\kb,\e)-g^2|\Delta|^2\right] ^{-1},
        \qquad
        \mathcal{F}_h\equiv\left[ \hat{G}_{0p}^{-1}(\kb,\e)\hat{G}_{0h}^{-1} (\kb,\e)-g^2|\Delta|^2\right] ^{-1},
        \\
        &
        \mathcal{G}_p
        \equiv \mathcal{F}_p \hat{G}_{0h}^{-1},
        \qquad
        \mathcal{G}_h
        \equiv 
        \mathcal{F}_h \hat{G}_{0p}^{-1}.
 \end{aligned}
 \end{align}
 $\mathcal{F}_{p/h}$ and   $\mathcal{G}_{p/h}$ have the same causality structure in the Keldysh space as $G_{0}$ (see Eq.~\ref{eq:causality}) and also satisfy the fluctuation-dissipation relation as in Eq.~\ref{eq:FDT-f}.
Specifically, their retarded/advanced components are given by
 \begin{align}
						\begin{aligned}\label{eq:mF}
							&\mathcal{F}_{p}^{(R/A)}(\kb,\e)	= \mathcal{F}_{h}^{(R/A)}(\kb,\e)
									=
								\left[		\left(\e  \pm i\eta \right)^2 - E_k^2 \right]^{-1},
                                \\
                                &\mathcal{G}_{p}^{(R/A)}(\kb,\e)	 
									=
								\frac{\e + (\frac{k^2}{2m}-\mu) }{\left(\e  \pm i\eta \right)^2 - E_k^2 },
                                \qquad
                                \mathcal{G}_{h}^{(R/A)}(\kb,\e)	 
									=
								\frac{\e - (\frac{k^2}{2m}-\mu)  }{\left(\e  \pm i\eta \right)^2 - E_k^2 },
						\end{aligned}
					\end{align}
                    where
                    \begin{align}
						\begin{aligned}
                                &E_{k}	= \sqrt{(\frac{k^2}{2m}-\mu)^2+g^2|\Delta|^2}.
						\end{aligned}
					\end{align}
Substituting Eq.~\ref{eq:mF} into the saddle point equation in Eq.~\ref{eq:SPEQ} leads to the gap equation
 \begin{align}
						\begin{aligned}
							&\varepsilon_0-2\mu
							=
							\frac{g^2}{2}
							\int_{\kb} 
							\frac{\tanh(E_k/2T) }{E_k}.
						\end{aligned} 
					\end{align} 

\subsection{Bosonic and fermionic self energies}

We then consider fluctuations beyond the saddle point at the Gaussian order 
\begin{align}
    \delta \phicl=\phicl-\Delta,
    \qquad
    \delta \phiq=\phiq.
\end{align}
By expanding $\Tr\ln G_{\phi}$ in terms of $\delta\phi$ up to quadratic order, $S_{\msf{eff}}$ reduces to
					\begin{align}\label{eq:S2}
						\begin{aligned}
							i S_2
							=&
							i \int
							\begin{bmatrix}
								\delta\bar{\phi}(\qb,\ww) 
								&
								\delta {\phi}(-\qb,-\ww) 
							\end{bmatrix}
							\begin{bmatrix}
								\frac{1}{2} D_0^{-1} (\qb,\ww)-\Pi(\qb,\ww) 
                                &
                                -\Pi_2(\qb,\ww)
                                \\
                                -\Pi_2(\qb,\ww)
                                &
                                \frac{1}{2} D_0^{-1} (\qb,\ww)-\Pi(\qb,\ww)
							\end{bmatrix}
							\begin{bmatrix}
								\delta {\phi}(\qb,\ww) 
                                \\
								\delta \bar{\phi}(-\qb,-\ww) 
							\end{bmatrix}
                            \\
                            &+\int \Tr \ln
							\begin{bmatrix}
								\hat{G}_{0p}^{-1} & -g\Delta 
								\\
								-g \bar{\Delta}  & \hat{G}_{0h}^{-1} 
							\end{bmatrix}.
						\end{aligned}
					\end{align}
 where the second term represents the saddle point action $S_{\msf{eff}}[\phicl=\Delta,\phiq=0]$.
$\Pi$ and $\Pi_2$ stand for the normal and anomalous bosonic self-energy. They acquire the same causality structure as $D_0^{-1}$ and also satisfy the fluctuation-dissipation relation as in Eq.~\ref{eq:FDT-b}. Their retarded components are given by the following integrals
 \begin{align}\label{eq:Pi0}
					\begin{aligned}
						i\Pi^{(R)}(\qb,\ww)
						=&
						\frac{g^2}{2}
						\int
                        \left[
						\mathcal{G}_{p}^{(K)}(\kb+\qb,\e+\ww) 
						\mathcal{G}_{h}^{(A)}(\kb,\e)  
						+
						\mathcal{G}_{p}^{(R)}(\kb+\qb,\e+\ww) 
						\mathcal{G}_{h}^{(K)}(\kb,\e) 
                        \right],
						\\
						i\Pi_2^{(R)}(\qb,\ww)
						=&
						\frac{g^4\Delta^2}{2}
						\int
						\mathcal{F}_{p}^{(K)}(\kb+\qb,\e+\ww) 
						\mathcal{F}_{p}^{(A)}(\kb,\e)  
                        +
						\mathcal{F}_{p}^{(R)}(\kb+\qb,\e+\ww) 
						\mathcal{F}_{p}^{(K)}(\kb,\e). 
					\end{aligned}
					\end{align}

 In this paper, we are interested in the temperature regime above the critical temperature $T>T_c$, where the order parameter vanishes $\Delta=0$. In this regime, with $\Delta=0$, it is straightforward to see from Eq.~\ref{eq:Gphi-2} and Eq.~\ref{eq:Pi0} that the anomalous bosonic self-energy vanishes $\Pi_2=0$, while the normal self-energy $\Pi$ now reduces to
\begin{align}\label{eq:Pi}
\begin{aligned}
i\Pi^{(R)}(\qb,\ww)
=&
\frac{g^2}{2}
\int
\left[ 
G_{0p}^{(K)}(\kb+\qb,\e+\ww) 
G_{0h}^{(A)} (\kb,\e) 
+
G_{0p}^{(R)}(\kb+\qb,\e+\ww) 
G_{0h}^{(K)} (\kb,\e) 
\right] 
\\
=&
-\frac{g^2}{2}
\int
\left[ 
G_{0p}^{(K)}(\kb+\qb,\e+\ww) 
G_{0p}^{(R)} (-\kb,-\e) 
+
G_{0p}^{(R)}(\kb+\qb,\e+\ww) 
G_{0p}^{(K)} (-\kb,-\e) 
\right].
\end{aligned}
\end{align}
Note that the bosonic self-energy $\Pi$ enters the interaction dressed bosonic Green's function through
\begin{align}
\begin{aligned}\label{eq:D}
    D(\qb,\ww)=&
    \left[ D_0^{-1}(\qb,\ww)-2 \Pi(\qb,\ww) \right]^{-1} .
\end{aligned}
\end{align}	

For the derivation of the leading order fermionic self-energy $\Sigma(\kb,\e)$, we can perform an expansion of $G_{\phi}$ in terms of $\phi$. From the quadratic term, we find
\begin{align}
\begin{aligned}
&
\Sigma (\kb,\e) 
=
g^2\int_{\qb,\ww}
\int \D(\bar{\phi},\phi)
e^{iS_{\phi}}
 \Phi(\qb,\ww)
G_{0}(\kb-\qb,\e-\ww) 
 \Phi(-\qb,-\ww),
\end{aligned}
\end{align}
where $S_{\phi}$ is the bare bosonic action (see Eq.~\ref{eq:Z1}).
This leads to
\begin{align}
\begin{aligned}
    \Sigma_{p}(\kb,\e)=g^2 \int_{\qb,\ww}
    \left[
    D^{(K)}_0(\qb,\ww)G_{0h}(\kb-\qb,\e-\ww) 
    +
    D^{(R)}_0(\qb,\ww)G_{0h}(\kb-\qb,\e-\ww) \tau_1
    +
    D^{(A)}_0(\qb,\ww)\tau_1G_{0h}(\kb-\qb,\e-\ww) 
    \right].
\end{aligned}
\end{align}
Using the causality structure of $G_0$ (Eq.~\ref{eq:causality}), we obtain
\begin{align}\label{eq:Sigma}
\begin{aligned}
    i\Sigma^{(R)}_{p} (\kb,\e)
    =\,&
    g^2
    \intl{\qb,\ww}
    \left\lbrace 
    D^{(K)}_0(\qb,\ww)G^{(A)}_{0p}(\qb-\kb,\ww-\e)
    +
    D^{(R)}_0(\qb,\ww)
    G^{(K)}_{0p}(\qb-\kb,\ww-\e)
    \right\rbrace.
\end{aligned}
\end{align}	
The advance and Keldysh components of $\Sigma_{p}$ can be determined by the causality structure and the fluctuation-dissipation relation Eq.~\ref{eq:FDT-f}, while the hole component of the fermionic self-energy $\Sigma_{h}$ is given by $\Sigma_{h}^{(R/A)}(\kb,\e)=-\Sigma_{p}^{(A/R)}(-\kb,-\e)$.

With the self-energy expression, one can find the dressed fermionic Green's function $G$ by
\begin{align}\label{eq:G}
\begin{aligned}
    G(\kb,\e)=&
    \left[ G_0^{-1}(\kb,\e)-\Sigma (\kb,\e) \right]^{-1}.
\end{aligned}
\end{align}	
Note that in Eqs.~\ref{eq:Pi} and~\ref{eq:Sigma}, we express the leading order bosonic and fermionic self-energies $\Pi$ and $\Sigma$ in terms of the integrals of noninteracting bosonic and fermionic Green's functions $D_0$ and $G_0$. The corresponding expressions for the self-consistent self-energies can be straightforwardly  obtained by replacing the noninteracting bosonic and fermionic Green's functions $D_0$ and $G_0$ in  Eqs.~\ref{eq:Pi} and~\ref{eq:Sigma} with the interaction dressed ones $D$ defined in Eq.~\ref{eq:D} and $G$ in Eq.~\ref{eq:G}.

\subsection{Particle number equation}

In the two channel model, the sum of the total number of open-channel fermions (including both spin species) $N_f$ and twice the  number of closed-channel bosons $2N_b$ is conserved
\begin{align}
\begin{aligned}
    N_f+2N_b=N.
\end{aligned}
\end{align}
This constraint leads to a particle number equation that determines the chemical potential $\mu$ in terms of the detuning $\e_0$. Together with the gap equation, the particle number equation also etermines the critical temperature $T_c$ as a function of detuning $\e_0$. In this section, we provide a detailed derivation of the particle number equation at and above the critical temperature $T \geq T_c$. This calculation is a generalization of Nozi\`eres-Schmitt-Rink theory~\citep{nozieres1985} for the two-channel model, and a similar derivation in Matsubara
formalism can be found in Ref.~\citep{gurarie2007resonantly}.

We introduce a source field $V_0(\rb,t)$ that couples to the classical components of fermionic and two times the bosonic densities, and the corresponding action before the Keldysh rotation acquires the form:
	\begin{align}
					\begin{aligned}
						S_s[V_0]=&
						 \sum_{a=\pm}\sum_{\sigma=\uparrow,\downarrow} \int_{\rb,t}  V_0(\rb,t)
						\bpsi^{a}_{\sigma} (\rb,t)\psi^{a}_{\sigma} (\rb,t)
						 +2 \sum_{a=\pm} \int_{\rb,t}
						  V_0(\rb,t)\bphi^{a} (\rb,t) \phi^{a} (\rb,t).
					\end{aligned}
				\end{align}		
After the Keldysh rotation, it transforms to
						\begin{align}
						\begin{aligned}
							S_s[V_0]=
							V_0  \int_{\rb,t}  
							\bar{\Psi}(\rb,t) \tau_1\otimes \sigma_3 \Psi(\rb,t)
							+4 V_0 
							\int_{\rb,t}   \bphi(\rb,t)  \phi(\rb,t).
						\end{aligned}
					\end{align}		
The total atomic density $n(\rb,t)$ can be obtained from
 \begin{align}
								\begin{aligned}
												n(\rb,t)= \frac{1}{2i}\frac{\partial}{\partial V_0(\rb,t)}	\ln Z[V_0]	\bigg\lvert_{V_0=0}.		
								\end{aligned}
							\end{align}	

After first adding the source-field action $S_s[V_0]$ to $S_2$ in Eq.~\ref{eq:S2} at $\Delta=0$ , we integrate out the quadratic $\phi$ fluctuations and find
 					\begin{align}
						\begin{aligned}
							\ln Z [V_0]
							=&
							-\Tr\ln
								\left( D_0^{-1} -2\Pi+ 4V_0\right)
                            + \Tr \ln \left(\hat{G}_{0}^{-1}+V_0\tau_1\otimes \sigma^3 \right).
						\end{aligned}
					\end{align}
Combining the two equations above, we arrive at    
\begin{align}
								\begin{aligned}
												n= &
						\frac{1}{2i}	\int_{\kb,\e}
                        \left(
						G_{0p}^{(K)}  (\kb,\e)
                        -
                        G_{0h}^{(K)} (\kb,\e)
                        \right)
						-
						\frac{1}{2i} \frac{\partial}{\partial \mu}
                        \int_{\qb,\ww} \ln D^{(K)} (\qb,\ww)
                        \\
                        =&
                        -\int_{\kb}
                        \tanh (\xi_k^{(F)}/2T)
						-
                        \int_{\qb,\ww} \frac{\partial \delta(\qb,\ww)}{\partial \mu}\coth(\ww/2T),
                        \end{aligned}
							\end{align}
 where we have defined
 		\begin{align}
			\begin{aligned}
				\delta(\qb,\ww)
				=&
	               \im
				\ln \left[\ww - (\frac{q^2}{4m}-2\mu +\e_0)  + i \eta-\Pi^{(R)} (\qb,\ww) \right].
			\end{aligned}
		\end{align}
 In the $g\rightarrow 0$ limit, we can ignore the fluctuation contribution encoded by the bosonic self-energy $\Pi$, and the particle number equation above reduces to
 		 \begin{align}
		 \begin{aligned}
		 n&=
		 2\int_{\kb}
		 n_F\left( \xi_k^{(F)} \right) 
		 +
		 2\int_{\qb}
		 n_B\left( \xi_q^{(B)}\right)
		 &=
		 \frac{1}{2^{1/2}\pi^{3/2}}
		 (mT)^{3/2}
		 \left[ 
		 -\Li_{3/2} \left( -\exp(\dfrac{\mu}{T})\right) 
		 +2^{3/2} \Li_{3/2} \left( \exp(\dfrac{2\mu- \e_0}{T})\right) 
		 \right],
		 \end{aligned}
		 \end{align}
		 where $n_{F/B}$ represents the Fermi-Dirac/Bose-Einstein distribution function and
     $\Li_s(z)\equiv \sum_{k=1}^{\infty}\frac{z^k}{k^s}$ denotes the Polylogarithm function.

 \section{Retarded correlation function of the stress tensor}

We are now ready to evaluate the retarded correlation function of the stress tensor $\GG^{(R)}_{T\alpha\beta}(0,\WW)$, above the critical temperature using Eq.~\ref{eq:GT-0}.
Note that the stress tensor $T_{xy}$ consists of a fermionic part $T_{xy}^{(F)}$ and a bosonic part $T_{xy}^{(B)}$.
We therefore decompose the retarded correlation function into fermionic ($\GG^{(R)}_{TFF}$), bosonic ($\GG^{(R)}_{TBB}$), and mixed components ($\GG^{(R)}_{TFB}$ and $\GG^{(R)}_{TBF}$):
					\begin{align}\label{eq:GT-ab}
						\begin{aligned}
							&
							i\GG^{(R)}_{T\alpha\beta}(0,\WW)
							=\,
							\braket{
								T_{xy}^{(\alpha)cl}(0,\WW)
								T_{xy}^{(\beta)q}(0,-\WW)
							}
                            =
                            \int 
						\D \left( \bar{\psi}, \psi \right) 
						\D \left( \bar{\phi}, \phi \right) 
						e^{
						iS_\psi+iS_\phi+iS_{\msf{int}}
						}
						T_{xy}^{(\alpha)cl}(0,\WW)
						T_{xy}^{(\beta)q}(0,-\WW),
                            \quad
                            \alpha,\beta=F,B.
						\end{aligned}
					\end{align}
Inserting the explicit expressions for the fermionic and bosonic noninteracting stress tensors $T_{xy}^{(F/B)}$ in Eq.~\ref{eq:Txy-Keldysh}, we obtain
                    \begin{subequations}\label{eq:GT-1}
					\begin{align}
						&\begin{aligned}\label{eq:GTFF}
							i\GG^{(R)}_{TFF}(0,\WW)
							=&
							\frac{1}{2}
							\int_{\kb_1,\kb_2,\e_1,\e_2}
							\frac{k_{1x} k_{1y}k_{2x} k_{2y}}{m^2}
                            \int \D(\bar{\phi},\phi)e^{iS_{\msf{eff}}[\bar{\phi},\phi]}
							\\
							&\times
							\left\lbrace
							\begin{aligned}
								&	\Tr
								\left[
								\hat{\tau}^1 \sigma^3 
								G_{\phi}
								(\kb_1,\e_1+\WW;\kb_2,\e_2)
								\sigma^3 
								G_{\phi}
								(\kb_2,\e_2-\WW;\kb_1,\e_1)
								\right]
								\\
								-&	\Tr
								\left[
								\hat{\tau}^1 \sigma^3 
								G_{\phi}
								(\kb_1,\e_1+\WW;\kb_1,\e_1)
								\right]
								\Tr
								\left[
								\sigma^3 
								G_{\phi}
								(\kb_2,\e_2-\WW;\kb_2,\e_2)
								\right]
							\end{aligned}
							\right\rbrace ,
                            \end{aligned}
							\\
                            &\begin{aligned}\label{eq:GTBB}
                            i\GG^{(R)}_{TBB}(0,\WW)
							=&\frac{1}{2}
							\int_{\qb_1,\ww_1,\qb_2,\ww_2}
							\frac{q_{1x} q_{1y}q_{2x} q_{2y}}{m^2}
                            \int \D(\bar{\phi},\phi)e^{iS_{\msf{eff}}[\bar{\phi},\phi]}
							\bar{\phi} (\qb_1,\ww_1)
							\phi(\qb_1,\ww_1+\WW)
							\bar{\phi} (\qb_2,\ww_2)
							\tau_1
							\phi(\qb_2,\ww_2-\WW),
                            \end{aligned}
                            \\
                            &\begin{aligned}\label{eq:GTFB}
							i\GG^{(R)}_{TFB}(0,\WW)
							=&
							-\frac{i}{2}
							\int_{\kb,\qb,\e,\ww}
							\frac{k_x k_yq_x q_y}{m^2}
                            \int \D(\bar{\phi},\phi)e^{iS_{\msf{eff}}[\bar{\phi},\phi]}
								\Tr
								\left( 
								\tau_1\sigma^3 
								G_{\phi}
								(\kb,\e+\WW;\kb,\e)
								\right)
								\bar{\phi} (\qb,\ww)
								\tau_1
								\phi(\qb,\ww-\WW)
								,
                            \end{aligned}
                            \\
                            &\begin{aligned}\label{eq:GTBF}
							i\GG^{(R)}_{TBF}(0,\WW)
							=&
                            -\frac{i}{2}
							\int_{\kb,\qb,\e,\ww}
							\frac{k_x k_yq_x q_y}{m^2}
                            \int \D(\bar{\phi},\phi)e^{iS_{\msf{eff}}[\bar{\phi},\phi]}
								\Tr
								\left( 
								\sigma^3 
								G_{\phi}
								(\kb,\e-\WW;\kb,\e)
								\right)
								\bar{\phi} (\qb,\ww)
								\phi(\qb,\ww+\WW).
						\end{aligned}
					\end{align} 
                    \end{subequations}
Here we have integrated out the fermions and worked with the effective theory of bosons with the action $S_{{\msf{eff}}}[\bar{\phi},\phi]$ in Eq.~\ref{eq:Seff}. $G_{\phi}$ is defined in Eq.~\ref{eq:Gphi} and it can be considered as the Green's function for free fermions in the presence of an external field $\phi$ which couples with $\psi$ via $S_{\msf{int}}$ (Eq.~\ref{eq:Z0}). 

\begin{figure}[t]
\begin{centering}
\includegraphics[width=0.45\columnwidth]{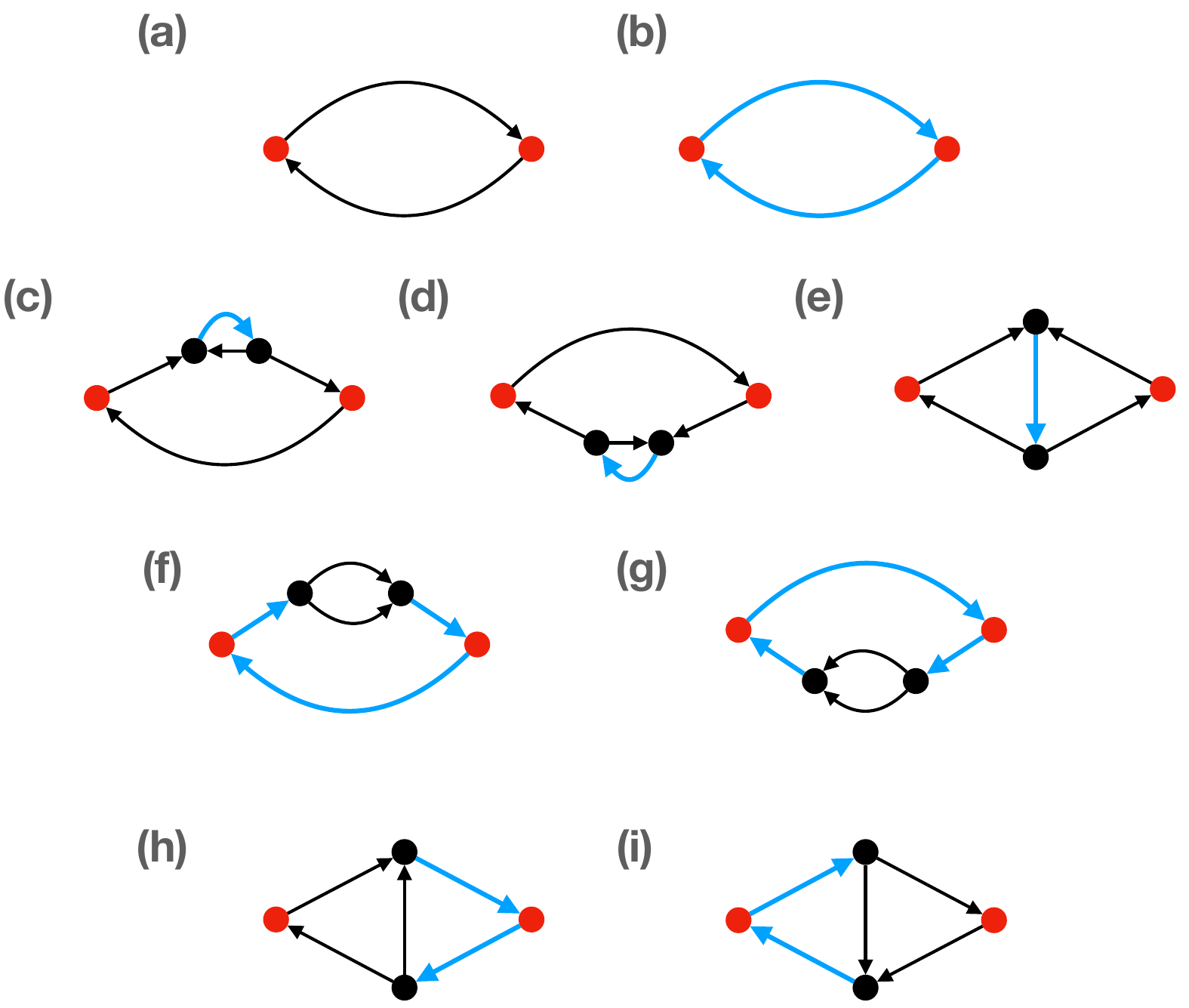} 
\par\end{centering}
\caption{
Diagrammatic expansion of the retarded correction function of the stress tensor in terms of small parameter $g$, shown up to order $O(g^2)$. Black (blue) lines denote the fermionic (bosonic) propagators, and black (red) vertices represent the fermion-boson coupling  (stress tensor) vertices.
The corresponding expressions for these diagrams, in their respective order, are given by Eqs.~\ref{eq:GTF-Drude}-\ref{eq:GTBF-1}.
}
\label{fig:Sup}
\end{figure}

We then perform an expansion in terms of the small parameter $g$ in the equation above, and obtain perturbatively the correlation function of the stress tensor $\GG_{T}^{(R)}$ up to order $g^2$. The corresponding diagrammatic expansion is summarized in Fig.~\ref{fig:Sup}.
At the zero order in $g$, the fermionic component of the stress tensor correlator $\GG^{(R)}_{TFF}$ receives the standard Drude contribution. This can be obtained from Eq.~\ref{eq:GTFF} by simply setting $G_{\phi}=G_0$. This fermionic Drude contribution is represented diagrammatically by Fig.~\ref{fig:Sup}(a), and its expression is given by 
	\begin{align}\label{eq:GTF-Drude}
						\begin{aligned}
							i\GG^{(R)}_{TFF-\msf{Drude}}(0,\WW)
							=\,&
							\int_{\kb,\e}
							\left(  \frac{k_x k_y}{m}\right)^2 
							\Tr
							\left( 
							{\tau}^1
								G_{0p} (\kb,\e+\WW) 
								G_{0p}(\kb,\e) 
							\right).
 			  	\end{aligned}
			  	\end{align}    
Note here the particle component of the noninteracting fermionic Green's function $G_{0p}$ is a matrix in the Keldysh space and has the causality structure in Eq.~\ref{eq:causality}.      

Similarly, the bosonic part of the stress tensor correlator $\GG^{(R)}_{TBB}$ exhibits a Drude contribution at the zero order, as shown in Fig.~\ref{fig:Sup}(b).  It is given by Eq.~\ref{eq:GTBB} after replacing $S_{\msf{eff}}$ with the bare bosonic action $S_{\phi}$ (Eq.~\ref{eq:Z0})     
			  	\begin{align}\label{eq:GTB-Drude}
			  	\begin{aligned}
			  	&i\GG^{(R)}_{TBB-\msf{Drude}}(0,\WW)
			  	=
			  	-\frac{1}{2}
			  	\int_{\qb,\ww}
			  	\left(  \frac{q_x q_y}{m}\right)^2 
			  	\Tr
			  	\left( 
			  	D_0(\qb,\ww+\WW)
			  	\tau_1
			  	D_0(\qb,\ww)
			  	\right).
			  	\end{aligned}
			  	\end{align} 
At the zeroth order, one can immediately see from Eqs.~\ref{eq:GTFB} and~\ref{eq:GTBF} that $\GG^{(R)}_{TFB}$ and $\GG^{(R)}_{TBF}$ vanish. Specifically,  at this order, the retarded correlation function of the stress tensor is just the sum of corresponding stress tensor correlators for noninteracting fermions and bosons. The associated Drude contributions are therefore divergent, and inclusion of interaction corrections is required to kill this divergence. See more detailed discussion in Sec.~\ref{sec:Drude}.

The interaction correction to the stress-tensor correlator first appears at the second order in $g$. One can see from Eq.~\ref{eq:GT-1} that,  the stress-tensor correlator vanishes at the first order in $g$, since it involves expectation values of terms linear in $\phi$ or $\bar{\phi}$ taken with respect to the effective action $S_{\msf{eff}}[\phi]$ with $\Delta=0$ for $T>T_c$. As a result, the leading order interaction corrections to the stress-tensor correlator appears at the order of $g^2$, and corresponding diagrams are listed in Fig.~\ref{fig:Sup}(c)-(i).

For the fermionic part, this includes the density of states (DOS) contributions shown in Fig.~\ref{fig:Sup}(c)(d) as well as the Maki-Thompson contribution shown in Fig.~\ref{fig:Sup}(e).  
Their explicit expressions are, in the respective order,
		  	\begin{subequations}
            \begin{align}
		  	&\begin{aligned}\label{eq:GTFF-DOS1}
		  	&i\GG^{(R)}_{TFF-\msf{DOS}1}(0,\Omega)=
		  	-ig^2\int_{\kb_1,\kb_2,\e_1,\e_2}
		  		\left( \frac{k_{1x} k_{1y}}{m}\right)^2 
		  		\\
		  		&\times\left\lbrace 
		  		\sum_{ab}
		  		\Tr
		  		\left[ 
		  		\hat{\tau}^1 
		  		G_{0p}(\kb_1,\e_1+\WW) 
		  		(\delta_{a1}\tau_1+\delta_{a2} ) 
		  		G_{0p}^T(\kb_2,\e_2) 
		  		(\delta_{b1}\tau_1+\delta_{b2} )
		  		G_{0p}(\kb_1,\e_1+\WW)
		  		G_{0p}(\kb_1,\e_1)
		  		\right] 
		  		D^{ab}_0(\kb_1+\kb_2,\e_1+\e_2+\WW)
		  		\right\rbrace ,
                \end{aligned}
		  	       \\
                 &  \begin{aligned}\label{eq:GTFF-DOS2}
		  		&i\GG^{(R)}_{TFF-\msf{DOS}2}(0,\WW)=
		  		-ig^2\int_{\kb_1,\kb_2,\e_1,\e_2}
		  		\left( \frac{k_{1x} k_{1y}}{m}\right)^2 
		  		\\
		  		&\times
		  		\left\lbrace 
		  		\sum_{ab}
		  		\Tr
		  		\left[ 
		  		\hat{\tau}^1 
		  		G_{0p}(\kb_1,\e_1+\WW)
		  		G_{0p}(\kb_1,\e_1) 
		  		(\delta_{a1}\hat{\tau}^1+\delta_{a2} ) 
		  		G_{0p}^T(\kb_2,\e_2) 
		  		(\delta_{b1}\hat{\tau}^1+\delta_{b2} ) 
		  		G_{0p}(\kb_1,\e_1)
		  		\right] 
		  		D^{ab}_0(\kb_1+\kb_2,\e_1+\e_2)
		  		\right\rbrace ,
                \end{aligned}
		  	\\
            &\begin{aligned}\label{eq:GTFF-MT}
            &i\GG^{(R)}_{TFF-\msf{MT}}(0,\WW)=
		  	-ig^2\int_{\kb_1,\kb_2,\e_1,\e_2}
		  	\left( \frac{k_{1x} k_{1y}k_{2x} k_{2y}}{m^2}\right) 
		  	\\
		  	&\times
		  	\left( 
		  	\sum_{ab}
		  	\Tr
		  	\left[ 
		  	\hat{\tau}^1 
		  	G_{0p}(\kb_1,\e_1+\WW) 
		  	(\delta_{a1}\tau_1+\delta_{a2} ) 
		  	G_{0p}^T(\kb_2,\e_2)
		  	G_{0p}^T(\kb_2,\e_2+\WW) 
		  	(\delta_{b1}\tau_1+\delta_{b2} ) 
		  	G_{0p}(\kb_1,\e_1)
		  	\right] 
		  	D^{ab}_0(\kb_1+\kb_2,\e_1+\e_2+\WW)
		  	\right) .
		  	\end{aligned}
		  	\end{align}
            \end{subequations}
The derivation of these components involves an expansion of $G_{\phi}$ in Eq.~\ref{eq:GTFF} in terms of $g$ (equivalently in terms of $\phi$ and $\bar{\phi}$): 
$G_{\phi} =G_0+gG_0 \Phi G_0+g^2G_0 \Phi G_0 \Phi G_0+O(g^3)$,  
and then a functional integration with respect to $S_{\msf{eff}}$. 
            
 At the same order of $g^2$, there are two additional diagrams contributing to the bosonic part of the stress-tensor correlator $\GG^{(R)}_{TBB}$, depicted in Fig.~\ref{fig:Sup}(f)(g). Similar to the DOS corrections in Fig.~\ref{fig:Sup}(c)(d), these are the leading self-energy corrections to the Drude contribution:
                \begin{subequations}
			  	\begin{align}
			  	&\begin{aligned}\label{eq:GTb-s1}
			  		&i\GG^{(R)}_{TBB-s1}(\Qb=\vex{0},\WW)
			  		=
			  		-\int_{\qb,\ww}
			  		\frac{q_{x}^2 q_{y}^2}{m^2}
			  		\Tr
			  		\left( 
			  		D_0(\qb,\ww+\WW)\Pi(\qb,\ww+\WW)D_0(\qb,\ww+\WW)
			  		\tau_1
			  		D_0(\qb,\ww)
			  		\right) 
                    \end{aligned},
                    \\
                 &   \begin{aligned}\label{eq:GTb-s2}
		  	&i\GG^{(R)}_{TBB-s2}(\Qb=\vex{0},\WW)
		  	=
		  	-\int_{\qb,\ww}
		  	\frac{q_{x}^2 q_{y}^2}{m^2}
		  	\Tr
		  	\left( 
		  	D_0(\qb,\ww+\WW)
		  	\tau_1
		  	D_0(\qb,\ww)\Pi(\qb,\ww)D_0(\qb,\ww)
		  	\right) .
			  	\end{aligned}
		  	\end{align} 
            \end{subequations}
Here in the derivation, we have used the expansion $D=(D_0^{-1}-2\Pi)^{-1}=D_0^{-1}+2D_0^{-1}\Pi D_0^{-1}+...$,
with $\Pi$ the bosonic self-energy.

For the boson-fermion mixed components to the stress tensor correlator $\GG^{(R)}_{TFB}$ and $\GG^{(R)}_{TBF}$, they become nonvanishing  at the quadratic order in $g$. Their leading order contributions are depicted in Fig.~\ref{fig:Sup}(h)(i), and take the following forms:
            \begin{subequations}\label{eq:GTBF-1}
		  	\begin{align}
		  	&\begin{aligned}\label{eq:GTFB-0}
            &i\GG^{(R)}_{TFB-0}(\vex{0},\WW)
		  	=
		  	-ig^2
		  	\int_{\kb,\qb,\e,\ww}
		  	\frac{k_x k_yq_x q_y}{m^2}
		  	\\
		  	&\times
		  	\left( 
		  	\sum_{ab}
		  	\Tr
		  	\left[ 
		  	\tau_1
		  	G_{0p}(\kb,\e+\WW)
		  	\left( 
		  	\delta_{a1}\tau_1 +\delta_{a2}
		  	\right) 	  
		  	G_{0p}^{T}(\qb-\kb,\ww-\e-\WW)
		  	\left( \delta_{b1}\tau_1 +\delta_{b2}
		  	\right) 
		  	G_{0p}(\kb,\e)
		  	\right] 
		  	\left( D_0(\qb,\ww) \tau_1 D_0(\qb,\ww-\WW)\right)^{ab} 
		  	\right) ,
            \end{aligned}
                \\
                &\begin{aligned}\label{eq:GTBF-0}
		  	&i\GG^{(R)}_{TBF-0}(\vex{0},\WW)
		  	=
		  	-ig^2
		  	\int_{\kb,\qb,\e,\ww}
		  	\frac{k_x k_yq_x q_y}{m^2}
		  			  	\\
		  	&\times
		  	\left( 
		  	\sum_{ab}
		  	\Tr
		  	\left[ 
		  	G_{0p}(\kb,\e-\WW)
		  	\left( 
		  	\delta_{a1}\tau_1 +\delta_{a2}
		  	\right) 
		  	G_{0p}^T(\qb-\kb,\ww-\e+\WW)
		  	\left( \delta_{b1}\tau_1 +\delta_{b2}
		  	\right) 
		  	G_{0p}(\kb,\e)
		  	\right] 
		  	\left( D_0(\qb,\ww)  D_0(\qb,\ww+\WW)\right)^{ab} 
		  	\right) .
		  	\end{aligned}
 		  	\end{align}
		  	\end{subequations}

 \section{Drude viscosity}~\label{sec:Drude}

 \subsection{Fermionic Drude viscosity}

 Using the causality structure for the fermionic Green's function~\ref{eq:causality} together with the Kubo formula Eq.~\ref{eq:Kubo}, we find that the contribution from $G_{TFF-\msf{Drude}}^{(R)}$ in Eq.~\ref{eq:GTF-Drude} to the shear viscosity, i.e., fermionic Drude contribution to the shear viscosity can be rewritten as the following integral
\begin{align}
\begin{aligned}\label{eq:F-Drude-0}
    \eta_{F}=\,&
    -
    \frac{1}{4T}
    \int_{\kb,\e}
    \left(  \frac{k_x k_y}{m}\right)^2 
    \sech^2 (\e/2T)
    \left[ G^{(R)}_{0p}(\kb,\e)-G^{(A)}_{0p}(\kb,\e) \right]^2.  
\end{aligned}
\end{align}
By substituting the noninteracting Green's function $G_0$ expression in Eq.~\ref{eq:G0-1}, it is straightforward to see that this integral diverges. This is not surprising as this zero order contribution is simply the viscosity for noninteracting fermions. 

This divergence can be removed by incorporating interaction self-energy corrections into the Durde viscosity diagram in Fig.~\ref{fig:Sup}(a). As a result, the noninteracting propagator $G_0$ there is replaced by the full interacting Green's function $G$. Diagrammatically, this is equivalent to resuming an infinite series of viscosity diagrams generated by self-energy insertions, including the second order DOS corrections depicted in Fig.~\ref{fig:Sup} (c)(d). After replacing $G_0$ with $G$ in Eq.~\ref{eq:F-Drude-0}, we obtain 
\begin{align}
\begin{aligned}\label{eq:F-Drude}
    \eta_{F}=\,&
    -
    \frac{1}{4T}
    \int_{\kb,\e}
    \left(  \frac{k_x k_y}{m}\right)^2 
    \sech^2 (\e/2T)
    \left[ G^{(R)}_{p}(\kb,\e)-G^{(A)}_{p}(\kb,\e) \right]^2 
    \\
    =\,&
    \frac{1}{T}
    \int_{\kb, \e}
    \left(  \frac{k_x k_y}{m}\right)^2 
    \sech^2 (\e/2T)
    \left[ 
    \frac{\im\Sigma^{(R)}_{p}(\kb,\e)}{\left( \e-\xi_k^{(F)}-\re \Sigma^{(R)}_{p}(\kb,\e)\right)^2 +\left( \im\Sigma^{(R)}_{p}(\kb,\e) \right)^2 }
    \right]^2.
\end{aligned}
\end{align}
We then ignore the effect from the real part of the fermionic self-energy which leads to the renormalization of atomic mass, and assume a small fermionic scattering rate:
	$
	\tau_F^{-1}(\kb,\e)=-2 \im \Sigma_p^{(R)}(\kb,\e)
	$
    due to small $g$. In this case,  the fermionic spectral function  can be considered as a narrow Lorenzian of width $1/2\tau_F$, which leads to
    \begin{align}\label{eq:F-Drude-t}
	\begin{aligned}
	\eta_{F}
	\approx &
	\frac{1}{2Tm^2}
	\int_{\kb}
	\left(  k_x k_y \right)^2 
	\sech^2 (\xi_k^{(F)}/2T)
	\tau_F(\kb,\xi_k^{(F)}).
	\end{aligned}
	\end{align}

    We note that the equation above also applies to the conventional Fermi liquid,  with different relaxation time $\tau_{F}$.
    Particularly, in the Fermi liquid case, the well-known Fermi liquid shear viscosity result by
    Abrikosov and Khalatnikov~\citep{Abrikosov1959} appears naturally from this integral.
	Note that the thermal factor $\sech^2(\xi_k^a/2T)$ in the integral for the fermionic Drude viscosity is sharply peaked at $\xi_k^{(F)}=0$. For a Fermi liquid, this restricts the momentum integration to momenta in the vicinity of the Fermi surface $k \approx k_F$, since the chemical potential lies close to the Fermi energy $\mu\approx E_F$. As a result, the momentum integral can be obtained by focusing on contributions around Fermi momentum $k_F$, which leads to
					\begin{align}
						\begin{aligned}
							\eta_{F}=\,&
                            \frac{	k_F^4}{m^2}
                            \frac{mk_F}{8\pi^3}
                            \tau_F(k_F,0)
    						\int_0^{2\pi} d\phi
    						\int_0^{\pi} d\theta \sin \theta (\sin \theta \cos \phi)^2 (\sin \theta \sin \phi)^2
							\int_{-\infty}^{\infty} d\e
							\frac{d}{d\e}\tanh(\e/2T)
							=\,
							\frac{1}{15\pi^2}
    						\frac{	k_F^5}{m}
    						\tau_F(k_F,0).
						\end{aligned}
					\end{align}
    This can be rewritten in the same form as Abrikosov and Khalatnikov's result~\citep{Abrikosov1959}, in terms of the particle density $n=k_F^3/3\pi^2$,
	\begin{align}\label{eq:AK}
	\eta_{AK}=\frac{1}{5}nmv_F^2 \tau_{F}(k_F,0).
	\end{align}

 \subsection{Bosonic Drude viscosity}

 In a way analogous to its fermionic counterpart, we find the bosonic Drude contribution to the shear viscosity from Eq.~\ref{eq:GTB-Drude}
\begin{align}\label{eq:B-Drude-0}
\begin{aligned}
    \eta_{B}=\,&
    -\frac{1}{8T}
    \int_{\qb,\ww}
    \left(  \frac{q_x q_y}{m}\right)^2 
    \csch^2(\ww/2T)
    \left[ D^{(R)}_0(\qb,\ww)-D^{(A)}_0(\qb,\ww) \right]^2.
\end{aligned}
\end{align}
As in the fermionic case, this integral diverges and requires the inclusion of the self-energy corrections to kill this divergence. This is equivalent to replace the noninteracting bosonic Green's function $D_0$ with the interacting dressed one $D$:
\begin{align}\label{eq:B-Drude}
\begin{aligned}
    \eta_{B}=\,&
    -\frac{1}{8T}
    \int_{\qb,\ww}
    \left(  \frac{q_x q_y}{m}\right)^2 
    \csch^2(\ww/2T)
    \left[ D^{(R)}(\qb,\ww)-D^{(A)}(\qb,\ww) \right]^2
    \\
    =&
    \frac{1}{8T}
    \int_{\qb,\ww}
    \left(  \frac{q_x q_y}{m}\right)^2 
    \csch^2(\ww/2T)
    \left[ 
    \frac{\im \Pi^{(R)}(\qb,\ww)}{\left(\ww-(\frac{q^2}{4m}-2\mu+\e_0)-\re \Pi^{(R)}(\qb,\ww)\right)^2+\left(\im \Pi^{(R)}(\qb,\ww)\right)^2}
    \right]^2,
\end{aligned}
\end{align} 
which for small $g$ further simplifies to
\begin{align}
	\begin{aligned}\label{eq:B-Drude-t}
	\eta_{B}
	\approx&
	\frac{1}{16Tm^2}
	\int_{\qb}
	(q_{x} q_{y})^2
	\csch^2 ({\xi}^{(B)}_{q}/2T)
	\tau_B(\qb,{\xi}^{(B)}_{q}).
	\end{aligned}
	\end{align}
Here, as in the fermionic Drude viscosity calculation, we have approximated the bosonic scattering function as a narrow Lorenzian with width given by bosonic scattering rate $\tau_B^{-1}(\qb,\ww)=-2 \im \Pi^{(R)}(\qb,\ww)$ and ignore the effect from the real part of the bosonic self-energy which leads to the normalization of the detuning $\e_0$.

 \section{Interaction correction to the shear viscosity}

In this section, we derive the interaction corrections to the shear viscosity that cannot be grouped into the self-energy corrections to the fermionic and bosonic Drude viscosities discussed in the previous section.
This includes the fermionic MT correction to the shear viscosity, which arises from $\GG_{TFF-\msf{MT}}^{(R)}$ in Eq.~\ref{eq:GTFF-MT}, as well as the interaction corrections from the mix fermion-boson stress tensor correlators $\GG_{TFB-0}^{(R)}$ and $\GG_{TBF-0}^{(R)}$ in Eq.~\ref{eq:GTBF-1}.
We note that the fermionic DOS corrections from $\GG_{TFF-\msf{DOS1/2}}^{(R)}$ as well as the bosonic leading order self-energy corrections from $\GG^{(R)}_{TBB-s1/2}$ are included in the Drude viscosities formulas in Eqs.~\ref{eq:F-Drude-t} and~\ref{eq:B-Drude-t}. So these will not be considered in this section. Additionally, similar to the Drude viscosities in Eqs.~\ref{eq:F-Drude-0} and~~\ref{eq:B-Drude-0}, divergence appears if we only consider the leading order interaction correction by using the bare fermionic and bosonic Green's functions $G_0$ and $D_0$ instead of dressed ones $G$ and $D$ in stress tensor correlation integrals. As a result, as in the Drude viscosities case, we replace the noninteracting fermionic/bosonic Green's function $G_0/D_0$ with the interaction dressed ones $G/D$ in Eqs.~\ref{eq:GTFF-MT} and ~\ref{eq:GTBF} to remove the divergence.
Specifically, we find the three additionally interaction corrections
\begin{align}
	&\begin{aligned}
		&
		\eta_{MT}
		=
        +i\frac{g^2}{2}
			\int_{\kb_1,\kb_2,\e_1,\e_2}
			\frac{k_{1x}k_{1y}}{m}\frac{k_{2x}k_{2y}}{m}
			\dot{F}(\e_1)		
			(G^{(R)}(\kb_1,\e_1)-G^{(A)}(\kb_1,\e_1))^2
            \\
            &\times
			\left[ 
            \begin{aligned}
			&G^{(R)}(\kb_2,\e_2)G^{(A)}(\kb_2,\e_2)
			(D^{(R)}(\kb_1+\kb_2,\e_1+\e_2)-D^{(A)}(\kb_1+\kb_2,\e_1+\e_2))B(\e_1+\e_2)
            \\
			&+
			G^{(R)}(\kb_2,\e_2) (G^{(R)}(\kb_2,\e_2)-G^{(A)}(\kb_2,\e_2)) 
			D^{(R)}(\kb_1+\kb_2,\e_1+\e_2)
            F(\e_2)
            \\
            &
			+
			(G^{(R)}(\kb_2,\e_2)-G^{(A)}(\kb_2,\e_2)) G^{(A)}(\kb_2,\e_2)
			D^{(A)}(\kb_1+\kb_2,\e_1+\e_2)
            F(\e_2)
            \end{aligned}
			\right] 
        \\
        &
		-i\frac{g^2}{2}
		\int_{\kb_1,\kb_2,\e_1,\e_2}
		\frac{k_{1x}k_{1y}}{m}\frac{k_{2x}k_{2y}}{m}	\dot{F}(\e_2)		
		\\
		&\times \left\lbrace 
		\begin{aligned} 
			&G^{(R)}(\kb_2,\e_2) G^{(A)}(\kb_2,\e_2)
            \\
            &\times
			\left[ 
			\left( 
			G^{(R)}(\kb_1,\e_1)
			G^{(R)}(\kb_1,\e_1)
			+
			G^{(A)}(\kb_1,\e_1)
			G^{(A)}(\kb_1,\e_1)
			\right) 
			\left( 
			D^{(R)}(\kb_1+\kb_2,\e_1+\e_2)
			-
			D^{(A)}(\kb_1+\kb_2,\e_1+\e_2)
			\right) 
			B	(\e_1+\e_2)		
            \right.
            \\
            &\left.\qquad
			+
			\left( 
			D^{(R)}(\kb_1+\kb_2,\e_1+\e_2)
			+
			D^{(A)}(\kb_1+\kb_2,\e_1+\e_2)
			\right) 
			\left( 
			G^{(R)}(\kb_1,\e_1)
			G^{(R)}(\kb_1,\e_1)
			-
			G^{(A)}(\kb_1,\e_1)
			G^{(A)}(\kb_1,\e_1)
            \right) 			
			F(\e_1)
			\right] 
			\\
			&
			+
			G^{(R)} (\kb_2,\e_2)G^{(R)}(\kb_2,\e_2)
            \\
            &\times
			\left[ 
			-
			G^{(A)}(\kb_1,\e_1)
			G^{(A)}(\kb_1,\e_1)
			\left( 
			D^{(R)}(\kb_1+\kb_2,\e_1+\e_2)
			-
			D^{(A)}(\kb_1+\kb_2,\e_1+\e_2)
			\right) 
			B	(\e_1+\e_2)		
            \right.
            \\
            &\qquad
            \left.
			-
			D^{(R)}(\kb_1+\kb_2,\e_1+\e_2)
			\left( 
			G^{(R)}(\kb_1,\e_1)
			G^{(R)}(\kb_1,\e_1)
			-
			G^{(A)}(\kb_1,\e_1)
			G^{(A)}(\kb_1,\e_1)
			\right) 	
			F(\e_1)
			\right] 
			\\
			&
			+
			G^{(A)}(\kb_2,\e_2) G^{(A)} (\kb_2,\e_2)
            \\
                        &\times
			\left[ 
			-
			G^{(R)}(\kb_1,\e_1)
			G^{(R)}(\kb_1,\e_1)
			\left( 
			D^{(R)}(\kb_1+\kb_2,\e_1+\e_2)
			-
			D^{(A)}(\kb_1+\kb_2,\e_1+\e_2)
			\right) 
			B	(\e_1+\e_2)		
            \right.
            \\
            &\qquad \left.
			-
			D^{(A)}(\kb_1+\kb_2,\e_1+\e_2)
			\left( 
			G^{(R)}(\kb_1,\e_1)
			G^{(R)}(\kb_1,\e_1)
			-
			G^{(A)}(\kb_1,\e_1)
			G^{(A)}(\kb_1,\e_1)
			\right) 	
			F(\e_1)
			\right] 
		\end{aligned}
		\right\rbrace,
	\end{aligned}
\\
&		\begin{aligned}
	&		\eta_{FB}
    =  +\frac{ig^2}{2}
    \int_{\kb_1,\kb_2,\e_1,\e_2}
    \frac{k_{1x} k_{1y}}{m}
    \frac{k_{2x} k_{2y}}{m}
    \dot{B}(\e_1)
    	(D^{(R)}(\kb_1,\e_1) -D^{(A)}(\kb_1,\e_1))^2
        \\
        &\times
    	\left[ 
        \begin{aligned}
    	&G^{(R)}(\kb_2,\e_2)
    	G^{(A)}(\kb_2,\e_2)
    	G^{(K)}(\kb_1-\kb_2,\e_1-\e_2)
        \\
    	+&
    	G^{(R)}(\kb_2,\e_2)
    	G^{(K)}(\kb_2,\e_2)
    	G^{(A)}(\kb_1-\kb_2,\e_1-\e_2)
        \\
    	+&
        G^{(K)}(\kb_2,\e_2)
    	G^{(A)}(\kb_2,\e_2)
    	G^{(R)}(\kb_1-\kb_2,\e_1-\e_2)
        \end{aligned}
    	\right] 
    	\\
    	&-\frac{ig^2}{2}
    \int_{\kb_1,\kb_2,\e_1,\e_2}
    \frac{k_{1x} k_{1y}}{m}
    \frac{k_{2x} k_{2y}}{m}
    	\dot{F}(\e_2) 
        \\
        &\times
    	\left\lbrace 
    	\begin{aligned}
    	&
    	{G}^{(R)}(\kb_2,\e_2){G}^{(A)}(\kb_2,\e_2)
        \\
        &\times
    	\left[ 
        \begin{aligned}
        &
    	(G^{(A)}(\kb_1-\kb_2,\e_1-\e_2)+G^{(R)}(\kb_1-\kb_2,\e_1-\e_2))
    	\left(D^{(R)}(\kb_1,\e_1) D^{(R)}(\kb_1,\e_1)-D^{(A)}(\kb_1,\e_1) D^{(A)}(\kb_1,\e_1)\right)B(\e_1)
        \\
                &
    	+G^{(K)}(\kb_1-\kb_2,\e_1-\e_2)
    	\left(D^{(R)}(\kb_1,\e_1) D^{(R)}(\kb_1,\e_1)+D^{(A)}(\kb_1,\e_1) D^{(A)}(\kb_1,\e_1)\right)
        \\
        \end{aligned}
    	\right] 
    	\\
    	&+
    	{G}^{(R)}(\kb_2,\e_2)
    	G^{(R)}(\kb_2,\e_2)
        \\
        &\times
    	\left[
        \begin{aligned}
        &-G^{(K)}(\kb_1-\kb_2,\e_1-\e_2)
    	D^{(R)}(\kb_1,\e_1) D^{(R)}(\kb_1,\e_1)
        \\
        &
    	-
    	G^{(A)}(\kb_1-\kb_2,\e_1-\e_2)
    	\left(D^{(R)}(\kb_1,\e_1) D^{(R)}(\kb_1,\e_1)-D^{(A)}(\kb_1,\e_1) D^{(A)}(\kb_1,\e_1)\right)B(\e_1)
        \end{aligned}
    	\right] 
    	\\
    	&+	
    	{G}^{(A)}(\kb_2,\e_2)
    	G^{(A)}(\kb_2,\e_2)
        \\
        &\times
    	\left[ 
        \begin{aligned}
    	&-G^{(K)}(\kb_1-\kb_2,\e_1-\e_2)
    	D^{(A)}(\kb_1,\e_1) D^{(A)}(\kb_1,\e_1)
        \\
    	&-
    	G^{(R)}(\kb_1-\kb_2,\e_1-\e_2)
    	\left(D^{(R)}(\kb_1,\e_1) D^{(R)}(\kb_1,\e_1)-D^{(A)}(\kb_1,\e_1) D^{(A)}(\kb_1,\e_1)\right)B(\e_1)
        \end{aligned}
    	\right] 
    	\end{aligned}
    	\right\rbrace ,
\end{aligned}
\\
	&	\begin{aligned}
	&		\eta_{BF}
    =  +\frac{ig^2}{2}
    \int_{\kb_1,\kb_2,\e_1,\e_2}
    \frac{k_{1x} k_{1y}}{m}
    \frac{k_{2x} k_{2y}}{m}
    \dot{F}(\e_1)
    	(G^{(R)}(\kb_1,\e_1) -G^{(A)}(\kb_1,\e_1))^2
        \\
        &\times
    	\left[ 
        \begin{aligned}
    	&D^{(R)}(\kb_2,\e_2)
    	D^{(A)}(\kb_2,\e_2)
    	G^{(K)}(\kb_2-\kb_1,\e_2-\e_1)
        \\
    	+&
    	D^{(R)}(\kb_2,\e_2)
    	D^{(K)}(\kb_2,\e_2)
    	G^{(R)}(\kb_2-\kb_1,\e_2-\e_1)
        \\
    	+&
        D^{(K)}(\kb_2,\e_2)
    	D^{(A)}(\kb_2,\e_2)
    	G^{(A)}(\kb_2-\kb_1,\e_2-\e_1)
        \end{aligned}
    	\right] 
    	\\
    	&-\frac{ig^2}{2}
    \int_{\kb_1,\kb_2,\e_1,\e_2}
    \frac{k_{1x} k_{1y}}{m}
    \frac{k_{2x} k_{2y}}{m}
    	\dot{B}(\e_2) 
        \\
        &\times
    	\left\lbrace 
    	\begin{aligned}
    	&
    	{D}^{(R)}(\kb_2,\e_2){D}^{(A)}(\kb_2,\e_2)
        \\
        &\times
    	\left[ 
        \begin{aligned}
        &
    	G^{(K)}(\kb_2-\kb_1,\e_2-\e_1)
    	\left(G^{(R)}(\kb_1,\e_1) G^{(R)}(\kb_1,\e_1)+G^{(A)}(\kb_1,\e_1) G^{(A)}(\kb_1,\e_1)\right)
        \\
        &
    	+(G^{(A)}(\kb_2-\kb_1,\e_2-\e_1)+G^{(R)}(\kb_2-\kb_1,\e_2-\e_1))
    	\left(G^{(R)}(\kb_1,\e_1) G^{(R)}(\kb_1,\e_1)-G^{(A)}(\kb_1,\e_1) G^{(A)}(\kb_1,\e_1)\right)F(\e_1)
        \end{aligned}
    	\right] 
    	\\
    	&+
    	{D}^{(R)}(\kb_2,\e_2)
    	D^{(R)}(\kb_2,\e_2)
        \\
        &\times
    	\left[
        \begin{aligned}
        &-G^{(K)}(\kb_1-\kb_2,\e_1-\e_2)
    	G^{(R)}(\kb_1,\e_1) G^{(R)}(\kb_1,\e_1)
        \\
        &
    	-
    	G^{(R)}(\kb_2-\kb_1,\e_2-\e_1)
    	\left(G^{(R)}(\kb_1,\e_1) G^{(R)}(\kb_1,\e_1)-G^{(A)}(\kb_1,\e_1) G^{(A)}(\kb_1,\e_1)\right)F(\e_1)
        \end{aligned}
    	\right] 
    	\\
    	&+	
    	{D}^{(A)}(\kb_2,\e_2)
    	D^{(A)}(\kb_2,\e_2)
        \\
        &\times
    	\left[ 
        \begin{aligned}
    	&-G^{(K)}(\kb_2-\kb_1,\e_2-\e_1)
    	G^{(A)}(\kb_1,\e_1) G^{(A)}(\kb_1,\e_1)
        \\
    	&-
    	G^{(A)}(\kb_2-\kb_1,\e_2-\e_1)
    	\left(G^{(R)}(\kb_1,\e_1) G^{(R)}(\kb_1,\e_1)-G^{(A)}(\kb_1,\e_1) G^{(A)}(\kb_1,\e_1)\right)F(\e_1)
        \end{aligned}
    	\right] 
    	\end{aligned}
    	\right\rbrace,
\end{aligned}
\end{align}
where we have defined $F(\e)=\tanh(\e/2T)$ and $B(\ww)=\coth(\ww/2T)$.

 Keeping only the leading order terms that are singular in $\tau_F^{-1},\tau_B^{-1}\rightarrow 0$, we can ignore terms involving products of two retarded or advanced Green's functions: $G^{(R)} G^{(R)}$, $G^{(A)} G^{(A)}$, $D^{(R)} D^{(R)}$, or $D^{(A)} D^{(A)}$. The expressions above then reduce to
 \begin{subequations}
\begin{align}
	&\begin{aligned}
			\eta_{MT}			=
		&-i\frac{g^2}{2T}
			\int_{\kb_1,\kb_2,\e_1,\e_2}
            (\frac{k_{1x}k_{1y}}{m}\frac{k_{2x}k_{2y}}{m})
			\sech^2(\e_1/2T)
			G^{(R)}(\kb_1,\e_1)
            G^{(A)}(\kb_1,\e_1)
            \\
            &
			\times G^{(R)}(\kb_2,\e_2)G^{(A)}(\kb_2,\e_2)
			\left[D^{(R)}(\kb_1+\kb_2,\e_1+\e_2)-D^{(A)}(\kb_1+\kb_2,\e_1+\e_2)\right]
            \left[
            \coth\left((\e_1+\e_2)/2T\right)
            -
            \tanh(\e_2/2T)
            \right].\label{eq:etamt}
	\end{aligned}	
    \\
		&\begin{aligned}
			\eta_{FB}
			=& +\frac{ig^2}{2T}
			\int_{\kb_1,\kb_2,\e_1,\e_2}
			\frac{k_{1x}k_{1y}}{m}\frac{k_{2x}k_{2y}}{m}
				\csch^2(\e_1/2T)
				D^{(R)}(\kb_1,\e_1)D^{(A)}(\kb_1,\e_1)
                \\
                \times&
				G^{(R)}(\kb_2,\e_2) G^{(A)}(\kb_2,\e_2)
                \left[G^{(R)}(\kb_1-\kb_2,\e_1-\e_2)-G^{(A)}(\kb_1-\kb_2,\e_1-\e_2)
                \right]  
                \left[
				\tanh\left((\e_1-\e_2)/2T\right)
                +
                \tanh\left(\e_2/2T\right)
                \right].
		\end{aligned}
\\
	&\begin{aligned}
		\eta_{BF}=&
		 -\frac{ig^2}{2T}
			\int_{\kb_1,\kb_2,\e_1,\e_2}
			\frac{k_{1x}k_{1y}}{m}\frac{k_{2x}k_{2y}}{m}
			\sech^2(\e_1/2T)
            G^{(R)}(\kb_1,\e_1)
            G^{(A)}(\kb_1,\e_1)
            \\
            &\times
			D^{(R)}(\kb_2,\e_2)D^{(A)}(\kb_2,\e_2)
			\left[
			G^{(R)}(\kb_2-\kb_1,\e_2-\e_1)			
			-
            G^{(A)}(\kb_2-\kb_1,\e_2-\e_1)	
            \right]
            \left[
            \tanh\left(
            (\e_2-\e_1)/2T
            \right)
            -
            \coth(\e_2/2T)
            \right].\label{eq:etabf}
		\end{aligned}
\end{align}
\end{subequations}
\section{Numerical calculation}

We now discuss the details of numerical calculation of vertex corrections.
We first note that we can reduce the integration order in Eqs.~(\ref{eq:etamt}-\ref{eq:etabf}). In particular,
consider the following integral that represents an integral over ${\bf k}_{2}$
in Eqs.~(\ref{eq:etamt}-\ref{eq:etabf}).

\begin{align*}
 & M_{i,j}({\bf k}')=\int\frac{d^{3}k}{\left(2\pi\right)^{3}}\frac{k_{i}k_{j}}{m}f(\left|{\bf k}\pm{\bf k}'\right|,k),
\end{align*}
where $f$ can be any function that depends only on $\left|{\bf k\pm k'}\right|$
and $k$ . We can alternatively represent $M$ as follows

\[
M_{i,j}({\bf k}')={\bf e}_{i}\cdot\int\frac{d^{3}k}{\left(2\pi\right)^{3}}\frac{{\bf k}\otimes{\bf k}}{m}f(\left|{\bf k}\pm{\bf k}'\right|,k)\cdot{\bf e}_{j},
\]
where ${\bf e}_{i}$ are unit vectors. After some simple algebra we
get for the tensor part

\begin{align*}
 & \int\frac{d^{3}k}{\left(2\pi\right)^{3}}\frac{{\bf k}\otimes{\bf k}}{m}f(\left|{\bf k}\pm{\bf k}'\right|,k)\\
 & =\int\frac{k^{4}dk}{\left(2\pi\right)^{2}}\sin\theta d\theta\frac{1}{m}\left(\tilde{{\bf e}}_{x}\otimes\tilde{{\bf e}}_{x}\frac{\sin^{2}\theta}{2}+\tilde{{\bf e}}_{y}\otimes\tilde{{\bf e}}_{y}\frac{\sin^{2}\theta}{2}+\tilde{{\bf e}}_{z}\otimes\tilde{{\bf e}}_{z}\cos^{2}\theta\right)f(\left|{\bf k}\pm{\bf k}'\right|,k),
\end{align*}
where we introduced a new basis $\tilde{{\bf e}}_{i}$ with $\tilde{{\bf e}}_{z}\parallel{\bf k}'$($\tilde{{\bf e}}_{z}={\bf k'}/k'$)
and $\theta$ is the angle between ${\bf k}$ and ${\bf k}'$. In
the following we will be interested in case when $i\neq j$. Therefore,
we can extract the isotropic and anisotropic components as follows\textbackslash{}

\begin{align*}
M_{i,j}({\bf k}') & ={\bf e}_{i}\cdot\left\{ \alpha\tilde{{\bf e}}_{x}\otimes\tilde{{\bf e}}_{x}+\alpha\tilde{{\bf e}}_{y}\otimes\tilde{{\bf e}}_{y}+\alpha\tilde{{\bf e}}_{z}\otimes\tilde{{\bf e}}_{z}+\left(\beta-\alpha\right)\tilde{{\bf e}}_{z}\otimes\tilde{{\bf e}}_{z}\right\} \cdot{\bf e}_{j}\\
 & =\alpha{\bf e}_{i}\cdot{\bf e}_{j}+\left(\beta-\alpha\right)\frac{k'_{i}k'_{j}}{k^{2}},
\end{align*}
where 
\begin{align*}
\alpha & =\int\frac{k^{4}dk}{\left(2\pi\right)^{2}}\sin\theta d\theta\frac{1}{m}\frac{\sin^{2}\theta}{2}f(\left|{\bf k}\pm{\bf k}'\right|,k)\\
\beta & =\int\frac{k^{4}dk}{\left(2\pi\right)^{2}}\sin\theta d\theta\frac{1}{m}\cos^{2}\theta f(\left|{\bf k}\pm{\bf k}'\right|,k).
\end{align*}
Finally, for $i\neq j$ we get:

\begin{equation}
M_{i,j}({\bf k}')=\frac{k'_{i}k'_{j}}{k^{2}}\int\frac{k^{4}dk}{\left(2\pi\right)^{2}}\sin\theta d\theta\left(\frac{3\cos^{2}\theta-1}{2m}\right)f(\left|{\bf k}\pm{\bf k}'\right|,k)\label{eq:Mij}
\end{equation}
In order to simplify Eqs.~ further, in the limit of weak coupling
we apply the following approximations $G^{(R)}(\kb,\e)G^{(A)}(\kb,\e)\approx2\pi\delta\left(\epsilon-\xi_{{\bf k}}^{f}\right)\tau^{F}_{{\bf k}},$ $G^{R}({\bf k},\epsilon)-G^{A}({\bf k},\epsilon)\approx-2\pi i\delta\left(\xi_{{\bf k}}^{f}-\epsilon\right)$
and analogously for bosons. With these approximations, the angular
integral ($\angle$(${\bf k}_{1}$,${\bf k}_{2}$)) can be taken analytically
reducing the integrals to numerically computable. 

Let us now demonstrate that within these approximations $\eta_{BF}=\eta_{FB}$.
Utilizing approximations above, we get:

\begin{align}
\eta_{FB} & =\frac{\pi g^{2}}{T}\int\frac{k_{1x}k_{1y}}{m}\frac{k_{2x}k_{2y}}{m}\text{csch}^{2}(\frac{\xi_{{\bf k}_{1}}^{b}}{2T})\tau_{{\bf k}_{2}}^{f}\tau_{{\bf k_{1}}}^{b}\delta\left(\xi_{{\bf k_{1}-k_{2}}}^{f}-\xi_{{\bf k_{1}}}^{b}+\xi_{{\bf k}_{2}}^{f}\right)(\tanh\left(\frac{\xi_{{\bf k_{1}}}^{b}-\xi_{{\bf k}_{2}}^{f}}{2T}\right)+\tanh\left(\frac{\xi_{{\bf k}_{2}}^{f}}{2T}\right))\\
\eta_{BF} & =-\frac{\pi g^{2}}{T}\int\frac{k_{1x}k_{1y}}{m}\frac{k_{2x}k_{2y}}{m}\text{sech}^{2}(\frac{\xi_{{\bf k}_{1}}^{f}}{2T})\tau_{{\bf k}_{1}}^{f}\tau_{{\bf k}_{2}}^{b}\delta\left(\xi_{{\bf k_{2}-k_{1}}}^{f}-\xi_{{\bf k}_{2}}^{b}+\xi_{{\bf k}_{1}}^{f}\right)(\tanh\left(\frac{\xi_{{\bf k}_{2}}^{b}-\xi_{{\bf k}_{1}}^{f}}{2T}\right)-\coth\left(\frac{\xi_{{\bf k}_{2}}^{b}}{2T}\right)).\label{eq:loop}
\end{align}

Re-labeling variables in one of these expressions $\bf{k_1}\leftrightarrow\bf{k_2}$, and using:

\[
\text{csch}^{2}(\frac{\xi_{{\bf k}_{1}}^{b}}{2T})(\tanh\left(\frac{\xi_{{\bf k_{1}}}^{b}-\xi_{{\bf k}_{2}}^{f}}{2T}\right)+\tanh\left(\frac{\xi_{{\bf k}_{2}}^{f}}{2T}\right))+\text{sech}^{2}(\frac{\xi_{{\bf k}_{2}}^{f}}{2T})(\tanh\left(\frac{\xi_{{\bf k}_{1}}^{b}-\xi_{{\bf k}_{2}}^{f}}{2T}\right)-\coth\left(\frac{\xi_{{\bf k}_{1}}^{b}}{2T}\right))=0,
\]
we find that $\eta_{FB}=\eta_{BF}$ within our approximations. We also note that as can be seen in e.g. Eq.~\eqref{eq:loop}, the overall $g^2$ prefactor can be canceled with one $g^{-2}$ in one of the lifetimes $\tau^{F/B}$, resulting the whole expression having the same order of magnitude as $\eta_{F/B}$.
 \bibliography{bibl}